\documentclass[a4paper,10pt]{article}
\usepackage[utf8]{inputenc}
\usepackage[english]{babel}
\usepackage{algorithmic,algorithm}
\usepackage{color,colortbl}
\usepackage[colorlinks,citecolor=blue,linkcolor=black]{hyperref}
\usepackage{amsmath,amsfonts,amssymb,amsthm,mathptmx,graphicx,latexsym}
\usepackage[a4paper,includeheadfoot,pdftex]{geometry}
\usepackage{float, subfig,longtable,lscape,multirow}
\input{./Qcircuit.tex}
\newtheorem{example}{Example}
\newtheorem{remark}{Remark}
\newcommand{\etal}{\emph{et al. }}

\title{Concise Quantum Associative Memories with Nonlinear Search Algorithm}

\author{J.-P. TCHAPET NJAFA, S.G. NANA ENGO \\
\small{Laboratory of Photonics, Department of Physics,} \\
\small{University of Ngaoundere, POB 454 Ngaoundere, Cameroon}
}

\begin{document}
\maketitle
\begin{abstract}
The model of quantum associative memories (QAM) we propose here consists in 
simplifying and generalizing that of Rigui Zhou \etal \cite{zhou2012} who uses 
the quantum matrix with the binary decision diagram put forth by David 
Rosenbaumand \cite{Rosenbaum2010} and the Abrams and Lloyd's nonlinear search 
algorithm \cite{Abrams1998}. Our model gives the possibility to retrieve one of 
the sought states in multi-values retrieving scheme when a measure on the first 
register is done in $\mathcal{O}(c-r)$ time complexity. It is better than 
Grover's algorithm and  its modified form which need 
$\mathcal{O}(\sqrt{\frac{2^n} {m}})$ steps when they are used as the retrieval 
algorithm. $n$ is the number of qubit of the first register and  $m$ the number 
of values $x$ for which $f(x)=1$.  As the nonlinearity makes the system highly 
susceptible to noise, an analysis of the influence of the single qubit noise 
channels on the Nonlinear Search Algorithm of our model of QAM, shows a fidelity 
of about $0.7$ whatever the number of qubits existing in the first register.
\end{abstract}

\section{Introduction}

Quantum Neural Networks are Artificial Neural Networks functioning according to 
quantum laws. One of the useful Neural Networks is the Associative Memory which 
is an important tool for pattern recognition, intelligent control and artificial 
intelligence. Ventura and Martinez have built a model of Quantum Associative 
Memory (QAM) where the stored patterns are considered as the basis states of the 
memory quantum state \cite{ventura2000}. They used a modified version of the 
well-known Grover's quantum search algorithm in an unsorted database as the 
retrieval algorithm. In order to overcome the limits of that model to only solve 
the completion problem by doing data retrieving from noisy data, Ezhov \etal 
have used an \emph {exclusive} method of quantum superposition and Grover's 
algorithm with distributed queries \cite{ezhov2000} . However, their model still 
produces non-negligible probability of irrelevant classification. We have 
recently put forth an improved model of QAM with distributed query that reduces 
the probability of this irrelevant classification \cite{tchapet2012}.

The NonLinear Search Algorithm (NLSA) is based on the fact that it has been 
suggested that under some circumstances, the superposition principle of quantum 
theory might be violated. In other words, sometime a quantum system might have 
temporal nonlinear evolution. Therefore, nonlinear quantum computer could solve 
NP-complete and even \#P problems in polynomial time that Abrams and Lloyd 
argued in 1998 in their nowadays classic paper \cite{Abrams1998}. We recall that 
the NLSA of Abrams and Lloyd uses the Weinberg’s prescription and they based 
their argument on a general property of nonlinear evolutions in Hilbert spaces. 
This nonlinear evolution is the non-conservation of scalar products between 
nonlinearly evolving solutions of a nonlinear Schrödinger equation. This effect 
is called a mobility phenomenon. In order to avoid the fact that the Weinberg's 
formalism imply faster than light transmission \cite{polchinsky}, Czachor 
\cite{Czachor98} has proposed another description, based on the Polchinski-type 
one. The state components evolution depend upon hyperbolic tangents. In the 
present paper, we follow the  Czachor's description. Meyer and Wong give in 
\cite{meyer} another reason which can justify the use of nonlinear formalism in 
quantum mechanics: 
\begin{quotation}
``[\ldots] An obvious question is whether a modest, physically motivated
nonlinearity can still produce a computational advantage. In particular,
consider Bose-Einstein condensates (BECs). [\ldots] In general, describing such
many-body systems is difficult because of the many interaction terms. But under
certain conditions, one can assume that only two-body contact interactions
contribute and the s-wave scattering length $a$ is much less than the
interparticle spacing. Then using mean field theory, one finds that the system
is approximately described by a nonlinear Schrödinger equation [\ldots]''
\end{quotation}
Therefore, Meyer and Wongthey have used Gross-Pitaevskii equation to build their 
nonlinear search algorithm.

Rigui \etal \cite{zhou2012} have recently proposed a model of Ventura's
associative memory which uses Binary Superposed Quantum Decision Diagram (BSQDD)
as a learning process. They also used the above nonlinear algorithm of Abrams
and Lloyd as a retrieving process for multi-values retrieval. Although the
learning process of their model is good, there is some ambiguities on how the
memory evolves and how the multi-values retrieval arises. First of all, there is
no exact description about operator $U_1$ which links the first register denoted
by $\ket{\psi_i}$ to the second register denoted by $\ket{\gamma}$. Secondly the
use of simple binary decision diagram to represent states $\ket{\Phi}$ (see 
step 2 on section 3.2) seems to
only show the way to attempt the needed state. Moreover the nonlinear operator
denoted by $U_3$ seems to be use on particular state, not on supposed state (see
step 3 on section 3.3 and section 4.2 in \cite{zhou2012}). There is also no
indication on how a measure will give a needed state according to the fact that
nonlinear search algorithm leaves the first register in a superposed state.

In this paper, as the primary innovation, we propose a concise NLSA for QAM 
with a method to retrieve one of the sought states,  especially in multi-values 
retrieving scheme when a measure is done only on the first register with 
$\mathcal{O}(c-r)$ time complexity. The parameters $c$ and $r$ are obtained as 
follow. If $n$ is the number of qubits of first register, $p\leq2^n$ the number 
of stored patterns and if the values of $t$ qubits are known (i.e., $t$ qubits 
have been measured or are already be disentangled to others, or the oracle acts 
on a subspace of $(n-t)$ qubits), then we have the number of stored patterns 
$q\leq p$ and the number of values $x$ for which $f(x)=1$ $m\leq q$.   If 
$c=\mathtt{ceil}(\log_2{q})$ is the least integer greater or equal to 
$\log_2{q}$ and $r=\mathtt{int}(\log_2{m})$ the integer part of $\log_2{m}$. 
Thus, our model simplifies and generalizes that of Rigui \etal \cite{zhou2012}.

However, if the strength of the nonlinearity provides a large computational 
advantage, it also makes the system highly susceptible to noise which appears as 
a true bottleneck that may limit the usefulness of the NLSA. As another  
innovation of this paper, we investigate the effects of noise in the algorithm 
by considering the bit-flip quantum channel modeling environmentally induced 
noise. We assume that at most a single complete bit-flip error occurs on one of 
the data qubits. It should be noted that the problem of the influence of noise 
on the Grover quantum search algorithm has been extensively studied by various 
researchers \cite{salas,gawron}.

The paper is organized as follows: section \ref{sec:Snl} clearly describes the 
NLSA proposed by Abrams and Lloyd. Section \ref{sec:Cal} presents the QAM with 
NLSA, hereafter noted QAM-NLSA, with a new method to retrieve one of the sought 
states in multi-values retrieving scheme. In section \ref{sec:noise}, we 
introduce the single qubit noise channels model to the NLSA and we analyze its 
influence on our model of QAM. At the end, a short conclusion is provided in 
section \ref{sec:Concl}. But we start with a short description of parameters 
used in this paper.

The following parameters will be used throughout the paper:
\begin{itemize}
 \item $n$ is the number of qubit of the first register,
 \item $\ket{z}$ is the initial state of the first register,
 \item $x$ is the needed value while $\ket{x}$ is the corresponding state,
 \item $p\leq2^n$ the number of stored patterns,
 \item $q\leq p$ the number of stored patterns if the values of $t$ qubits are 
known,
\item $c=\mathtt{ceil}(\log_2{q})$, i.e. the least integer greater or equal to
$\log_2{q}$,
\item $m\leq q$ the number of values $x$ for which $f(x)=1$,
\item $r=\mathtt{int}(\log_2{m})$ is the integer part of $\log_2{m}$.
\end{itemize}
 
\section{Nonlinear search algorithm}
\label{sec:Snl}

Suppose there is a unitary transformation $\mathtt{U}_f$: the oracle or the
\emph{black box} which acts as follows: for a set of inputs between $0$ and
$2^n-1$, there is at most one $x$ for which $f(x)=1$ and the other values give
$0$. Let us consider two registers; the first register which is an $n$-qubit
system is to compute inputs and the second which is a single-qubit system is
to compute the answer of the oracle. We can define the function $f$ as
\begin{equation}
 \begin{split}
  f: & \mathcal{H}^{\otimes n}\longrightarrow \mathcal{H}\\
    & \ket{y}\longmapsto\ket{\delta_{xy}}
 \end{split}
\end{equation}
where $\mathcal{H}^{\otimes n}$ is a Hilbert space of $2^n$ dimensions.

The nonlinear algorithm of Abrams and Lloyd aims to disentangle the flag 
qubit from the first register as a 
measure on the flag qubit can tell us if there is at most a value $x$ for which 
$f(x)=1$. This is done by transforming the part of the flag qubit that is 
$\ket{0}$ to $\ket{1}$. They claim that it is not possible to do by using 
linear operators of quantum information. The Abrams and Lloyd nonlinear 
algorithm is summarized by the Algorithm \ref{alg:algo1}.

\begin{algorithm}
\caption{Nonlinear search algorithm (NLSA)}
\begin{algorithmic}[1]
\STATE Put the first register in the superposed state of all the $N$ values
and the flag qubit to $\ket{0}$
\STATE Apply the oracle $\mathtt{U}_f$
\FOR{each qubit of the first register with the flag qubit}
\STATE\label{alg:Balgo13} Apply the unitary operator $\mathtt{U}$
\STATE\label{alg:algo14} 
\begin{enumerate}
\item\label{alg:algo141} apply the nonlinear operator 
$\mathtt{NL}^-$
\item\label{alg:algo142} apply the nonlinear operator 
$\mathtt{NL}^+$
\end{enumerate}
\STATE\label{alg:Ealgo13} Apply the Hadamard operator $\mathtt{W}$ on the qubit 
of the first register and the NOT operator $\mathtt{X}$ on the flag qubit
\ENDFOR
\STATE Observe the flag qubit
\end{algorithmic}
	\label{alg:algo1}
\end{algorithm}

Let $\ket{\psi}$ be the state which describes all the system and assume that
all $N=2^n$ inputs are computed in the first register with equal amplitude:
\begin{equation}
 \ket{\psi}=\frac{1}{\sqrt{N}}\sum_y^N\ket{y}\ket{0}.
 \label{eq:or0}
\end{equation}
Applying the oracle yields 
\begin{equation}
 \begin{split}
  \mathtt{U}_f\ket{\psi}&=\frac{1}{\sqrt{N}}\sum_y^N\ket{y}\ket{f(y)}\\
&=\frac{1}{\sqrt{N}}\left(\sum_{y\neq x}^N\ket{y}\ket{0}+\ket{x}\ket{1}\right).
 \end{split}
 \label{eq:oracle}
\end{equation}

To describe the disentanglement algorithm, we consider the binary forms of 
values and assume that there is at most one value $x$ which gives $f(x)=1$. Let 
$\ket{j_nj_{n-1}\hdots j_1}$  and $\ket{i_ni_{n-1}\hdots i_1}$ be the binary 
forms of states $\ket{y}$ and $\ket{x}$ respectively, with $j_k,i_k\in\{0,1\}$. 
Equations (\ref{eq:or0}) and (\ref{eq:oracle}) can be rewritten as
\begin{equation}
 \ket{\psi}=\frac{1}{\sqrt{2^n}}\left[\sum_{\substack{j_nj_{n-1}\hdots j_1=0 \\
j_nj_{n-1}\hdots j_1\neq i_ni_{n-1}\hdots i_1}}^1\ket{j_nj_{n-1}\hdots j_1}
+\ket{i_ni_{n-1}\hdots i_1}\right]\ket{0},
\end{equation}
and
\begin{equation}
 \mathtt{U}_f\ket{\psi}=\ket{\Psi}=\frac{1}{\sqrt{2^n}}\left[\sum_{\substack{
j_nj_{n-1}\hdots j_1=0 \\j_nj_{n-1}\hdots j_1\neq i_ni_{n-1}\hdots i_1}}^1
\ket{j_nj_{n-1}\hdots j_1}\ket{0} +\ket{i_ni_{n-1}\hdots i_1}\ket{1} \right].
\label{eq:poraracle}
\end{equation}
Highlighting the least significant qubit (LSQ) of the first register, equation
(\ref{eq:poraracle}) can be helpfully written as
\begin{equation}
\ket{\Psi}=\frac{1}{\sqrt{2^n}}\left[\sum_{\substack{j_nj_{n-1}\hdots j_1=0\\
j_nj_{n-1}\hdots j_2\neq i_ni_{n-1}\hdots i_2}}^1\ket{j_nj_{n-1}\hdots j_1}
\ket{0}+\ket{i_ni_{n-1}\hdots(1-i_1)}\ket{0}+\ket{i_ni_{n-1}\hdots i_1}\ket{1}
\right].
\label{eq:equaO}
\end{equation}
The state (\ref{eq:equaO}) must be viewed as the general binary form of the 
system after the action of the oracle. It summarizes all particular states 
given by Czachor in \cite{Czachor98} and removes ambiguities given by his 
notation. Indeed, his equation
\begin{equation}
 \frac{1}{\sqrt{2^n}}\sum_{j_nj_{n-1}\hdots j_2=0}^1[\ket{j_nj_{n-1}\hdots 0_1}
\ket{1}+\ket{j_nj_{n-1}\hdots1_1}\ket{0}],
\end{equation}
suggests that there is $2^{n-1}$ values $x$ for which $f(x)=1$, and not $s=1$ 
as he claims.

Considering the subsystem of only the LSQ of the first register $\ket{\ell}$ and
the flag qubit $\ket{k}$, $k,\ell\in\{0,1\}$, the computer will be in one of
the following states where we ignore the normalization constants,
\begin{subequations}
 \begin{align}
  &\ket{00}+\ket{10},  \label{eq:NL}\\
  &\ket{10}+\ket{01},  \label{eq:NLO}\\
  &\ket{00}+\ket{11}.  \label{eq:NLT}
 \end{align}
 \label{eq:NLe}
\end{subequations}
The left part of the equation (\ref{eq:equaO}) suggests that the state 
(\ref{eq:NL}) occurs with the highest probability whereas the state 
$\ket{01}+\ket{11}$ does not appear because the variable $x$ is supposed to be 
unique.

The nonlinear evolution (NLE), step \ref{alg:Balgo13} to step \ref{alg:Ealgo13}
of Algorithm \ref{alg:algo1}, aims to transform the states (\ref{eq:NLO}) and
(\ref{eq:NLT}) to $\ket{01}+\ket{11}$ while leaving the state (\ref{eq:NL})
unchanged. The NLE part of the algorithm then acts as follows:

\begin{itemize}
 \item[\textbf{Step \ref{alg:Balgo13}.}] Apply the 2-qubit operator
 \begin{equation}
  \mathtt{U}=\frac{1}{\sqrt{2}}\begin{pmatrix}
                       1 & 0 & 0 & 1\\
                       0 & 1 & 1 & 0\\
                       0 & 1 & -1 & 0\\
                       1 & 0 & 0 & -1
                      \end{pmatrix}
 \end{equation}
on the states (\ref{eq:NLe}):
\begin{subequations}
 \begin{align}
 & \mathtt{U}(\ket{00}+\ket{10})=\frac{1}{\sqrt{2}}(\ket{00}+\ket{01}
 -\ket{10}+\ket{11}),\\
 & \mathtt{U}(\ket{10}+\ket{01})=\sqrt{2}\ket{01},\\
 & \mathtt{U}(\ket{00}+\ket{11})=\sqrt{2}\ket{00}.
 \end{align}
\end{subequations}
\item[\textbf{Step \ref{alg:algo14}.\ref{alg:algo141}.}]  Apply the nonlinear
1-qubit operator $\mathtt{NL}^-$ on the flag qubit:
\begin{subequations}
  \begin{align}
 & \mathtt{NL}^-[\frac{1}{\sqrt{2}}(\ket{00}+\ket{01}-\ket{10}+\ket{11})]
=\sqrt{2}\ket{0}(\alpha\ket{0}+\beta\ket{1}),\label{eq:equan-0}\\
 & \mathtt{NL}^-(\sqrt{2}\ket{01})=\sqrt{2}\ket{00},   \label{eq:equan-1}\\
 & \mathtt{NL}^-(\sqrt{2}\ket{00})=\sqrt{2}\ket{00}.   \label{eq:equan-2}
\end{align}
\end{subequations}
where $\alpha,\beta\in\mathbb{C},\,|\alpha|^2+|\beta|^2=1$. As we see on the
state (\ref{eq:equan-0}), the action of the 1-qubit nonlinear operator
$\mathtt{NL}^-$ on the state $\frac{1}{\sqrt{2}}(\ket{0}\pm\ket{1})$ is not
specified. This gives some flexibility to choose the nonlinear gate
$\mathtt{NL}^-$\cite{Abrams1998}. On the states (\ref{eq:equan-1}) and
(\ref{eq:equan-2}), the operator $\mathtt{NL}^-$ maps the two flag qubits
$\ket{0}$ and $\ket{1}$ to the state $\ket{0}$. Thus, it must be seen as the NOT
gate $\mathtt{X}$ in case of the state (\ref{eq:equan-1}) and the identity gate
$\mathbb{I}$ in case of the state (\ref{eq:equan-2}).

\item[\textbf{Step \ref{alg:algo14}.\ref{alg:algo142}.}] Apply the second
nonlinear 1-qubit operator $\mathtt{NL}^+$ on the flag qubit:
\begin{subequations}
  \begin{align}
 &\mathtt{NL}^+[\sqrt{2}\ket{0}(\alpha\ket{0}+\beta\ket{1})]=\sqrt{2}\ket{01},
\label{eq:equaT}\\
 &  \mathtt{NL}^+(\sqrt{2}\ket{00})=\sqrt{2}\ket{00}.
\end{align}
\end{subequations}
The nonlinear operator $\mathtt{NL}^+$ acts as the identity gate $\mathbb{I}$ on
the state $\ket{0}$. The general form of the unitary matrix $\mathtt{NL}^+$
which maps the generic 1-qubit $\alpha\ket{0}+\beta\ket{1}$ to $\ket{1}$ is
\begin{equation}
 \mathtt{M}=\begin{pmatrix}
    \mp\gamma\beta & \pm\gamma\alpha\\
    \alpha^{\ast} & \beta^{\ast}
   \end{pmatrix},\,\,\gamma=\pm i\text{ or }\pm 1\;(i^2=-1),\label{eq:matri}
\end{equation}
where $\,\alpha,\,\beta\in\mathbb{C},\,|\alpha|^2+|\beta|^2=1$.

It is noteworthy the matrix (\ref{eq:matri}) corrects that of  Rigui \etal 
\cite{zhou2012} who claim that matrix  $\mathtt{M}$ must be
\begin{equation}
 \mathtt{V}=\begin{pmatrix}
    1 & \frac{1}{\beta}\\
    0 & -\frac{\alpha}{\beta}
   \end{pmatrix},
\end{equation}
which is unfortunately not a unitary matrix like matrix (\ref{eq:matri}) as
required by quantum information processing. Furthermore, the matrix $\mathtt{V}$
 yields a wrong result
\begin{equation}
 \mathtt{V}[\sqrt{2}\ket{0}(\alpha\ket{0}+\beta\ket{1})]=\sqrt{2}\ket{0}[
(\alpha+1)\ket{0}-\alpha\ket{1}],
\end{equation}
and not $\sqrt{2}\ket{01}$ as expected.

\item[\textbf{Step \ref{alg:Ealgo13}.}] Apply the NOT gate $\mathtt{X}$ on the
flag qubit and the Hadamard gate $\mathtt{W}$ on the first qubit.
\end{itemize}

We summarize below the nonlinear evolution of the states of equations 
(\ref{eq:NLe}) and give their corresponding circuits:
\begin{subequations}
  \begin{equation}
\ket{00}+\ket{10}\xrightarrow{\mathtt{U}}\frac{1}{\sqrt{2}}(\ket{00}+\ket{01}
-\ket{10} +\ket{11})\xrightarrow{\mathtt{NL}^-}\sqrt{2}
\ket{0}(\alpha\ket{0}+\beta\ket{1})\xrightarrow{\mathtt{NL}^+}\sqrt{2}\ket{01}
\xrightarrow{\mathtt{W}\otimes\mathtt{X}}\ket{00}+\ket{10}.
\label{eq:e1}
 \end{equation}
 \begin{figure}[H]
\[  \Qcircuit @C=1.6em @R=2.em{
 &\dstick{\begin{matrix}\ket{00}\\+\ket{10}\end{matrix}}&\qw&\multigate{1}{
\mathtt{U}} &\qw&\qw&\qw&\multigate{1}{\mathtt{NL}^-}&\qw&\qw&\qw&\qw&\qw
&\gate{\mathtt{W}}
&\rstick{\ket{0}+\ket{1}}\qw\\
&&\qw&\ghost{\mathtt{U}}&\qw&\ustick{\begin{matrix}\frac{1}{\sqrt{2}}(\ket{00}
+\ket{01}\\-\ket{10}+\ket{11})\end{matrix}}\qw&\qw&\ghost{\mathtt{NL}^-}&\qw&
\ustick{\begin{matrix}\sqrt{2}\ket{0}(\alpha\ket{0}\\+\beta\ket{1})\end{matrix}
}\qw&\qw&\gate{\mathtt{NL}^+\equiv\mathtt{M}}&
\ustick{\sqrt{2}\ket{01}}\qw&\gate{\mathtt{X}}&\rstick{\ket{0}}\qw
  }
\]
  \caption{Equivalent circuit of the nonlinear evolution (\ref{eq:e1}).}
 \end{figure}
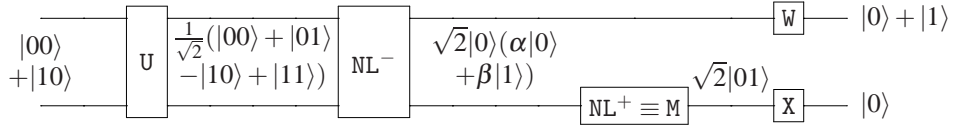

 \begin{equation}
\ket{10}+\ket{01}\xrightarrow{\mathtt{U}}\sqrt{2}\ket{01}\xrightarrow{\mathtt{NL
}^-}\sqrt{2}\ket{00}\xrightarrow{\mathtt{NL}^+}\sqrt{2}\ket{00}\xrightarrow{
\mathtt{W}\otimes\mathtt{X}}\ket{01}+\ket{11}.
\label{eq:e2}
 \end{equation}
  \begin{figure}[H]
  \centering
\[  \Qcircuit @C=1.6em @R=1.5em{
  &\dstick{\begin{matrix}\ket{10}\\+\ket{01}\end{matrix}}&\qw &\multigate{1}
{\mathtt{U}}&\qw&\qw&\qw&\qw&\qw&\gate{\mathtt{W}}
&\rstick{\ket{0}+\ket{1}}\qw\\
&&\qw&\ghost{\mathtt{U}}&\ustick{\sqrt{2}\ket{01}}
\qw&\gate{\mathtt{NL}^-\equiv\mathtt{X}}&\ustick{\sqrt{2}\ket{00}}\qw&\gate{
\mathtt{NL}^+\equiv\mathbb{I}}&\ustick{\sqrt{2}\ket{00}}\qw&\gate{\mathtt{X}}
&\rstick{\ket{1}}\qw}
\]
  \caption{Equivalent circuit of the nonlinear evolution (\ref{eq:e2}).}
 \end{figure}
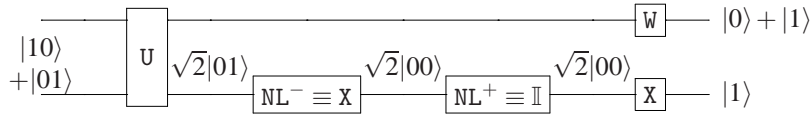
 \begin{equation}
\ket{00}+\ket{11}\xrightarrow{\mathtt{U}}\sqrt{2}\ket{00}\xrightarrow{\mathtt{NL
}^-}\sqrt{2}\ket{00}\xrightarrow{\mathtt{NL}^+}\sqrt{2}\ket{00}
\xrightarrow{\mathtt{W}\otimes\mathtt{X}}\ket{01}+\ket{11}.
\label{eq:e3}
 \end{equation}
  \begin{figure}[H]
  \centering
\[    \Qcircuit @C=1.6em @R=1.5em{
  &\dstick{\begin{matrix}\ket{00}\\+\ket{11}\end{matrix}}&\qw&\multigate{1}
{\mathtt{U}}&\qw&\qw&\qw&\qw&\qw&\gate{\mathtt{W}}
&\rstick{\ket{0}+\ket{1}}\qw\\
&&\qw&\ghost{\mathtt{U}}&\ustick{\sqrt{2}\ket{00}}
\qw&\gate{\mathtt{NL}^-\equiv\mathbb{I}}&\ustick{\sqrt{2}\ket{00}}\qw&
\gate{\mathtt{NL}^+\equiv\mathbb{I}}&\ustick{\sqrt{2}\ket{00}}\qw&\gate
{\mathtt{X}}&\rstick{\ket{1}}\qw
}
\]
  \caption{Equivalent circuit of the nonlinear evolution (\ref{eq:e3}).}
 \end{figure}
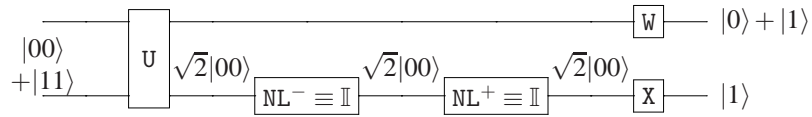
\end{subequations}

\begin{example}\label{exam1}
\begin{subequations}
For a better understanding, let us consider a simple case of an 4-qubit register
in the superposition states of all the $16$ possible values, plus a flag qubit.
The marked state is $\ket{2}=\ket{0010}$. We start with
\begin{multline}
 \ket{\psi}=\frac{1}{4}(\ket{0000}+\ket{0001}+\ket{0010}+\ket{0011}+\ket{0100
}+\ket{0101}+\ket{0110}+\ket{0111}\\+\ket{1000}+\ket{1001}+\ket{1010}+\ket{1011}
+\ket{1100} +\ket{1101}+\ket{1110}+\ket{1111})\ket{0}.
\label{eq:start_exam1}
\end{multline}
 The action of the oracle operator $\mathtt{U}_f$ yields
 \begin{multline}
 \ket{\Psi}=\frac{1}{4}(\ket{0000}\ket{0}+\ket{0001}\ket{0}+\ket{0010}\ket{
1}+\ket{0011}\ket{0}+\ket{0100}\ket{0}+\ket{0101}\ket{0}+\ket{0110}\ket{0}
\\+\ket { 0111}\ket{0}+\ket{1000}\ket{0}+\ket{1001}\ket{0}+\ket{1010}\ket{
0}+\ket{1011}\ket{0}+\ket{1100}\ket{0}+\ket{1101}\ket{0}\\+\ket{1110}\ket{0}
+\ket{1111}\ket{0}).
\end{multline}
\end{subequations}
Now we will describe the process as in \cite{Czachor98}, but with details on how 
the system is when the \texttt{NLE} gate is applied. $\mathtt{W_r}$ acts on the 
significant qubit and $\mathtt{X_f}$ acts on flag qubit. 

\begin{subequations}
We start by looking on the LSQ of the register
\begin{equation}
\begin{split}
 \ket{\Psi}=\frac{1}{4}&(
\ket{000{\boxed{{\color{blue}0}}}}\ket{{\boxed{{\color{blue}0}}}}+\ket{000{
\boxed{{\color{blue}1}}}}\ket{\boxed{{\color{blue}0}}}
+\ket{001{\boxed{{\color{blue}0}}}}\ket{{\boxed{{\color{blue}1}}}}
+\ket{001{\boxed{{\color{blue}1}}}}\ket{{\boxed{{\color {blue}0}}}} 
+\ket{010{\boxed{{\color{blue}0}}}}\ket{{\boxed{{\color{blue}0}}}}+\ket{010{
\boxed{{\color{blue}1} }}}\ket{{ \boxed{{\color{blue}0}}}} \\
&+\ket{011{\boxed{{\color{blue}0}}}}\ket{{\boxed{{\color{blue}0}}}}+\ket{011{
\boxed{{\color{blue}1} }}}\ket{{ \boxed{{\color{blue}0}}}} 
+\ket{100{\boxed{{\color{blue}0}}}}\ket{{\boxed{{\color{blue}0}}}}+\ket{100{
\boxed{{\color{blue}1} }}}\ket{{ \boxed{{\color{blue}0}}}} 
+\ket{101{\boxed{{\color{blue}0}}}}\ket{{\boxed{{\color{blue}0}}}}+\ket{101{
\boxed{{\color{blue}1} }}}\ket{{ \boxed{{\color{blue}0}}}} \\
&+\ket{110{\boxed{{\color{blue}0}}}}\ket{{\boxed{{\color{blue}0}}}}+\ket{110{
\boxed{{\color{blue}1} }}}\ket{{ \boxed{{\color{blue}0}}}}
+\ket{111{\boxed{{\color{blue}0}}}}\ket{{\boxed{{\color{blue}0}}}}+\ket{111{
\boxed{{\color{blue}1} }}}\ket{{ \boxed{{\color{blue}0}}}} )\\
=\frac{1}{4}&\left[(\ket{000}+\ket{010}+\ket{011}+\ket{100}+\ket{101}+\ket{110}
+\ket { 111})(\ket{00}+\ket{10})+\ket{001}(\ket{01}+\ket{10})\right]
 \end{split}
\end{equation}
Applying the \texttt{NLE} gate  the first time produces
\begin{equation}
\begin{split} 
\ket{\Psi_1}=\frac{1}{4}&(\mathtt{W_r}\mathtt{X_f})\mathtt{NL^+}\mathtt{NL^-}
\mathtt{U } \left [ (\ket { 000 } 
+\ket{010}+\ket{011}+\ket{100}+\ket{101}+\ket{110}
+\ket { 111})(\ket{00}+\ket{10})+\ket{001}(\ket{10}+\ket{01})\right]\\
=\frac{1}{4}&(\mathtt{W_r}\mathtt{X_f})\mathtt{NL^+}\mathtt{NL^-}\left[(\ket{000
}
+\ket{010}+\ket{011}+\ket{100}+\ket{101}+\ket{110}
+\ket { 111})(\frac{1}{\sqrt{2}}(\ket{00}+\ket{01}
 -\ket{10}+\ket{11})+\ket{001}(\sqrt{2}\ket{01})\right]\\
 =\frac{1}{4}&(\mathtt{W_r}\mathtt{X_f})\mathtt{NL^+}\left[(\ket{000}
+\ket{010}+\ket{011}+\ket{100}+\ket{101}+\ket{110}
+\ket{111})(\sqrt{2}\ket{0}(\alpha\ket{0}+\beta\ket{1}))+\ket{001}(\sqrt{2}\ket{
00} )\right]\\
=\frac{1}{4}&(\mathtt{W_r}\mathtt{X_f})\left[(\ket{000}
+\ket{010}+\ket{011}+\ket{100}+\ket{101}+\ket{110}
+\ket{111})(\sqrt{2}\ket{00})+\ket{001}(\sqrt{2}\ket{
00} )\right]\\
 =\frac{1}{4}&(
\ket{000{\boxed{{\color{blue}0}}}}\ket{{\boxed{{\color{blue}0}}}}+\ket{000{
\boxed{{\color{blue}1}}} }\ket{\boxed{{\color{blue}0}}}
+\ket{001{\boxed{{\color{blue}0}}}}\ket{{\boxed{{\color{blue}1}}}}
+\ket{001{\boxed{{\color{blue}1}}}}\ket{{\boxed{{\color{blue}1}}}} 
+\ket{010{\boxed{{\color{blue}0}}}}\ket{{\boxed{{\color{blue}0}}}}+\ket{010{
\boxed{{\color{blue}1} }}}\ket{{ \boxed{{\color{blue}0}}}} \\
&+\ket{011{\boxed{{\color{blue}0}}}}\ket{{\boxed{{\color{blue}0}}}}+\ket{011{
\boxed{{\color{blue}1} }}}\ket{{ \boxed{{\color{blue}0}}}}
+\ket{100{\boxed{{\color{blue}0}}}}\ket{{\boxed{{\color{blue}0}}}}+\ket{100{
\boxed{{\color{blue}1} }}}\ket{{ \boxed{{\color{blue}0}}}}
+\ket{101{\boxed{{\color{blue}0}}}}\ket{{\boxed{{\color{blue}0}}}}+\ket{101{
\boxed{{\color{blue}1} }}}\ket{{ \boxed{{\color{blue}0}}}}\\
&+\ket{110{\boxed{{\color{blue}0}}}}\ket{{\boxed{{\color{blue}0}}}}+\ket{110{
\boxed{{\color{blue}1} }}}\ket{{ \boxed{{\color{blue}0}}}}
+\ket{111{\boxed{{\color{blue}0}}}}\ket{{\boxed{{\color{blue}0}}}}+\ket{111{
\boxed{{\color{blue}1} }}}\ket{{ \boxed{{\color{blue}0}}}} )
\end{split}
\end{equation}

 Next looking on the second LSQ
\begin{equation}
\begin{split}
 \ket{\Psi_1}=\frac{1}{4}&(\ket{00{\boxed{{\color{blue}0}}}0}\ket{{\boxed{{
\color{blue}0}}}}+\ket{00{\boxed{{\color{blue}1}}}0}\ket{{
\boxed{{\color{blue}1}}}}+\ket{00{\boxed{{\color{blue}0}}}1}
\ket{{\boxed{{\color{blue}0}}}}+\ket{00{ \boxed{{\color{blue}1}}
}1}\ket{{\boxed{{\color{blue}1}}}} +\ket{01{
\boxed{{\color{blue}0}}}1}\ket{{\boxed{{\color{blue}0}}}}+\ket{01{\boxed{{
\color{blue}1}}}1}\ket{{\boxed{{\color{blue}0}}}}  \\ &+\ket{01{
\boxed{{\color{blue}0}}}0}\ket{{\boxed{{\color{blue}0}}}}+\ket{01{\boxed{{\color
{blue}1}}}0}\ket{{\boxed{{\color{blue}0}}}}+\ket{10{
\boxed{{\color{blue}0}}}1}\ket{{\boxed{{\color{blue}0}}}}+\ket{10{\boxed{{
\color{blue}1}}}1}\ket{{\boxed{{\color{blue}0}}}} +\ket{10{
\boxed{{\color{blue}0}}}0}\ket{{\boxed{{\color{blue}0}}}}+\ket{10{\boxed{{\color
{blue}1}}}0}\ket{{\boxed{{\color{blue}0}}}}\\ &+\ket{11{
\boxed{{\color{blue}0}}}1}\ket{{\boxed{{\color{blue}0}}}}+\ket{11{\boxed{{
\color{blue}1}}}1}\ket{{\boxed{{\color{blue}0}}}}+\ket{11{
\boxed{{\color{blue}0}}}0}\ket{{\boxed{{\color{blue}0}}}}+\ket{11{\boxed{{\color
{blue}1}}}0}\ket{{\boxed{{\color{blue}0}}}})\\
=\frac{1}{4}&\left[(\ket{010}+\ket{011}+\ket{100}+\ket{101}+\ket{110}
+\ket { 111})(\ket{00}+\ket{10})+(\ket{000}+\ket{001})(\ket{00}+\ket{11})\right]
 \end{split}
\end{equation}
Applying the \texttt{NLE} gate  the second time produces
\begin{equation}
\begin{split}
\ket{\Psi_2}=\frac{1}{4}&(\mathtt{W_r}\mathtt{X_f})\mathtt{NL^+}\mathtt{NL^-}
\mathtt{U }\left[(\ket{010}+\ket{011}+\ket{100}+\ket{101}+\ket{110
}+\ket{111})(\ket{00}+\ket{10})+(\ket{000}+\ket{001})(\ket{00}+\ket{11})\right]
\\
=\frac{1}{4}&(\mathtt{W_r}\mathtt{X_f})\mathtt{NL^+}\mathtt{NL^-}
\left[(\ket{010}+\ket{011}+\ket{100}+\ket{101}+\ket{110
}+\ket{111})(\frac{1}{\sqrt{2}}(\ket{00}+\ket{01}
 -\ket{10}+\ket{11})+(\ket{000}+\ket{001})(\sqrt{2}\ket{00})\right]
\\
=\frac{1}{4}&(\mathtt{W_r}\mathtt{X_f})\mathtt{NL^+}
\left[(\ket{010}+\ket{011}+\ket{100}+\ket{101}+\ket{110
}+\ket{111})(\sqrt{2}\ket{0}(\alpha\ket{0}+\beta\ket{1}))+(\ket{000}+\ket{001}
)(\sqrt{2}\ket{00})\right]
\\
=\frac{1}{4}&(\mathtt{W_r}\mathtt{X_f})
\left[(\ket{010}+\ket{011}+\ket{100}+\ket{101}+\ket{110
}+\ket{111})(\sqrt{2}\ket{00})+(\ket{000}+\ket{001}
)(\sqrt{2}\ket{00})\right]
\\
 =\frac{1}{4}&(\ket{00{\boxed{{\color{blue}0}}}0}\ket{{\boxed{{
\color{blue}1}}}}+\ket{00{\boxed{{\color{blue}1}}}0}\ket{{\boxed{{\color{blue}1}
}}}+\ket{00{\boxed{{\color{blue}0}}}1}\ket{{\boxed{{\color{blue}1}}}}+\ket{00{
\boxed{{ \color{blue}1}}}1}\ket{{\boxed{{\color{blue}1}}}}+\ket{01{
\boxed{{\color{blue}0}}}1}\ket{{\boxed{{\color{blue}0}}}}+\ket{01{\boxed{{
\color{blue}1}}}1}\ket{{\boxed{{\color{blue}0}}}}  \\ &+\ket{01{
\boxed{{\color{blue}0}}}0}\ket{{\boxed{{\color{blue}0}}}}+\ket{01{\boxed{{\color
{blue}1}}}0}\ket{{\boxed{{\color{blue}0}}}}+\ket{10{
\boxed{{\color{blue}0}}}1}\ket{{\boxed{{\color{blue}0}}}}+\ket{10{\boxed{{
\color{blue}1}}}1}\ket{{\boxed{{\color{blue}0}}}} +\ket{10{
\boxed{{\color{blue}0}}}0}\ket{{\boxed{{\color{blue}0}}}}+\ket{10{\boxed{{\color
{blue}1}}}0}\ket{{\boxed{{\color{blue}0}}}}\\ &+\ket{11{
\boxed{{\color{blue}0}}}1}\ket{{\boxed{{\color{blue}0}}}}+\ket{11{\boxed{{
\color{blue}1}}}1}\ket{{\boxed{{\color{blue}0}}}}+\ket{11{
\boxed{{\color{blue}0}}}0}\ket{{\boxed{{\color{blue}0}}}}+\ket{11{\boxed{{\color
{blue}1}}}0}\ket{{\boxed{{\color{blue}0}}}})
 \end{split}
\end{equation}

Now looking on the third qubit
\begin{equation}
\begin{split}
 \ket{\Psi_2}=\frac{1}{4}&(\ket{0{
\boxed{{\color{blue}0}}}00}\ket{{\boxed{{\color{blue}1}}}}+\ket{0{\boxed{{\color
{blue}1}}}00}\ket{{\boxed{{ \color{blue}0}} }}+\ket{0{
\boxed{{\color{blue}0}}} 01}\ket{{
\boxed{{\color{blue}1}}}}+\ket{0{\boxed{{\color{blue}1}}}01}\ket{{\boxed{{\color
{blue}0}}}}+\ket{0{ \boxed{{\color{blue}0}}} 10}\ket{{
\boxed{{\color{blue}1}}}}+\ket{0{\boxed{{\color{blue}1}}}10}\ket{{\boxed{{\color
{blue}0}}}}\\ &+\ket{0{ \boxed{{\color{blue}0}}} 11}\ket{{
\boxed{{\color{blue}1}}}}+\ket{0{\boxed{{\color{blue}1}}}11}\ket{{\boxed{{\color
{blue}0}}}}+\ket{1{
\boxed{{\color{blue}0}}}01}\ket{{\boxed{{\color{blue}0}}}}+\ket{1{\boxed{{\color
{blue}1}}}01}\ket{{\boxed{{\color{blue}0}}}} +\ket{1{
\boxed{{\color{blue}0}}}11}
\ket{{\boxed{{\color{blue}0}}}}+\ket{1{\boxed{{\color{blue}1}}}11}\ket{{\boxed{{
\color{blue}0}}}} \\ &+\ket{1{
\boxed{{\color{blue}0}}}00}\ket{{\boxed{{\color{blue}0}}}}+\ket{1{\boxed{{\color
{blue}1}}}00}\ket{{\boxed{{\color{blue}0}}}}+\ket{1{
\boxed{{\color{blue}0}}}10}\ket{{\boxed{{\color{blue}0}}}}+\ket{1{\boxed{{\color
{blue}1}}}10}\ket{{\boxed{{\color{blue}0}}}})\\
=\frac{1}{4}&\left[(\ket{000}+\ket{001}+\ket{010}+\ket{011})(\ket{10}+\ket{01}
)+(\ket{100}+\ket{101}+\ket{110}+\ket{111})(\ket{00}+\ket {10} )\right]
 \end{split}
\end{equation}
Applying the \texttt{NLE} gate  the third time  produces
\begin{equation}
\begin{split}
\ket{\Psi_3}=\frac{1}{4}&(\mathtt{W_r}\mathtt{X_f})\mathtt{NL^+}\mathtt{NL^-}
\mathtt{U}\left[(\ket{000}+\ket{001}+\ket{010}+\ket{011})(\ket{10}+\ket{01}
)+(\ket{100}+\ket{101}+\ket{110}+\ket{111})(\ket{00}+\ket{10})\right]\\
=\frac{1}{4}&(\mathtt{W_r}\mathtt{X_f})\mathtt{NL^+}\mathtt{NL^-}
\left[(\ket{000}+\ket{001}+\ket{010}+\ket{011})(\sqrt{2}\ket{01}
)+(\ket{100}+\ket{101}+\ket{110}+\ket{111})(\sqrt{2}(\ket{00}+\ket{01}
 -\ket{10}+\ket{11}))\right]\\
 =\frac{1}{4}&(\mathtt{W_r}\mathtt{X_f})\mathtt{NL^+}
\left[(\ket{000}+\ket{001}+\ket{010}+\ket{011})(\sqrt{2}\ket{00}
)+(\ket{100}+\ket{101}+\ket{110}+\ket{111})(\sqrt{2}\ket{0}(\alpha\ket{0}
+\beta\ket{1}))\right]\\
=\frac{1}{4}&(\mathtt{W_r}\mathtt{X_f})
\left[(\ket{000}+\ket{001}+\ket{010}+\ket{011})(\sqrt{2}\ket{00}
)+(\ket{100}+\ket{101}+\ket{110}+\ket{111})(\sqrt{2}\ket{0}(\sqrt{2}\ket{00}
)\right]\\
 =\frac{1}{4}&(\ket{0{
\boxed{{\color{blue}0}}}00}\ket{{\boxed{{\color{blue}1}}}}+\ket{0{\boxed{{\color
{blue}1}}}00}\ket{{\boxed{{ \color{blue}1}}}}+\ket{0{
\boxed{{\color{blue}0}}}10}\ket{{
\boxed{{\color{blue}1}}}}+\ket{0{\boxed{{\color{blue}1}}}10}\ket{{\boxed{{\color
{blue}1}}}}+\ket{0{ \boxed{{\color{blue}0}}}01}\ket{{
\boxed{{\color{blue}1}}}}+\ket{0{\boxed{{\color{blue}1}}}01}\ket{{\boxed{{\color
{blue}1}}}}\\ &+\ket{0{ \boxed{{\color{blue}0}}}11}\ket{{
\boxed{{\color{blue}1}}}}+\ket{0{\boxed{{\color{blue}1}}}11}\ket{{\boxed{{\color
{blue}1}}}}+\ket{1{
\boxed{{\color{blue}0}}}01}\ket{{\boxed{{\color{blue}0}}}}+\ket{1{\boxed{{\color
{blue}1}}}01}\ket{{\boxed{{\color{blue}0}}}} +\ket{1{
\boxed{{\color{blue}0}}}11}
\ket{{\boxed{{\color{blue}0}}}}+\ket{1{\boxed{{\color{blue}1}}}11}\ket{{\boxed{{
\color{blue}0}}}} \\ &+\ket{1{
\boxed{{\color{blue}0}}}00}\ket{{\boxed{{\color{blue}0}}}}+\ket{1{\boxed{{\color
{blue}1}}}00}\ket{{\boxed{{\color{blue}0}}}}+\ket{1{
\boxed{{\color{blue}0}}}10}\ket{{\boxed{{\color{blue}0}}}}+\ket{1{\boxed{{\color
{blue}1}}}10}\ket{{\boxed{{\color{blue}0}}}})
 \end{split}
\end{equation}

Finally, we look on the most significant qubit
\begin{equation}
\begin{split}
 \ket{\Psi_3}=\frac{1}{4}&(\ket{{\boxed{{\color{blue}0}}}000}\ket{{\boxed{{
\color{blue}1}}}}+\ket{{\boxed{{\color{blue}1}}}000}\ket{{\boxed{{\color{blue}0}
}}}+\ket{{\boxed{{\color{blue}0}}}001}\ket{{\boxed{{\color{blue}1}}}}+\ket{{
\boxed{{\color{blue}1}}}001}\ket{{ \boxed{{\color 
{blue}0}}}}+\ket{{\boxed{{\color{blue}0}}} 010}\ket{{
\boxed{{\color{blue}1}}}}+\ket{{\boxed{{\color{blue}1}}}010}\ket{{\boxed{{\color
{blue}0}}}}\\
&+\ket{{\boxed{{\color{blue}0}}}011}\ket{{\boxed{{\color{blue}1}}}}+\ket{{
\boxed {{\color{blue}1}}}011}\ket{{ 
\boxed{{\color{blue}0}}}}+\ket{{\boxed{{\color{blue}0}}} 100}\ket{{
\boxed{{\color{blue}1}}}}+\ket{{\boxed{{\color{blue}1}}}100}\ket{{\boxed{{\color
{blue}0}}}}+\ket{{\boxed{{\color{blue}0}}} 101}\ket{{
\boxed{{\color{blue}1}}}}+\ket{{\boxed{{\color{blue}1}}}101}\ket{{\boxed{{\color
{blue}0}}}}\\ &+\ket{{\boxed{{\color{blue}0}}} 110}\ket{{
\boxed{{\color{blue}1}}}}+\ket{{\boxed{{\color{blue}1}}}110}\ket{{\boxed{{\color
{blue}0}}}}+\ket{{ \boxed{{\color{blue}0}}} 111}\ket{{
\boxed{{\color{blue}1}}}}+\ket{{\boxed{{\color{blue}1}}}111}\ket{{\boxed{{\color
{blue}0}}}})\\
=\frac{1}{4}&\left[(\ket{000}+\ket{001}+\ket{010}+\ket{011}+\ket{100}+\ket{101}
+\ket{110}+\ket{111})(\ket{10}+\ket{01})\right]
\end{split}
\end{equation}
Applying the \texttt{NLE} gate  the last time  produces
\begin{equation}
\begin{split}
\ket{\Psi_4}=\frac{1}{4}&(\mathtt{W_r}\mathtt{X_f})\mathtt{NL^+}\mathtt{NL^-}
\mathtt{U}\left[(\ket{000}+\ket{001}+\ket{010}+\ket{011}+\ket{100
}+\ket{101}+\ket{110}+\ket{111})(\ket{10}+\ket{01})\right]\\
=\frac{1}{4}&(\mathtt{W_r}\mathtt{X_f})\mathtt{NL^+}\mathtt{NL^-}
\left[(\ket{000}+\ket{001}+\ket{010}+\ket{011}+\ket{100}+\ket{101}+\ket{110}
+\ket{111})(\sqrt{2}\ket{01})\right]\\
=\frac{1}{4}&(\mathtt{W_r}\mathtt{X_f})\mathtt{NL^+}\left[(\ket{000}+\ket{001}
+\ket{010}+\ket{011}+\ket{100}+\ket{101}+\ket{110}+\ket{111})(\sqrt{2}\ket{00}
)\right]\\
=\frac{1}{4}&(\mathtt{W_r}\mathtt{X_f})\left[(\ket{000}+\ket{001}+\ket{010}+\ket
{011}+\ket{100}+\ket{101}+\ket{110}+\ket{111})(\sqrt{2}\ket{00})\right]\\
=\frac{1}{4}&(\ket{{\boxed{{\color{blue}0}}}000}+\ket{{\boxed{{\color{blue}1}}}
000}+\ket{{ \boxed{{\color{blue}0}}}001}+\ket{{\boxed{{\color{blue}1}}}001}
+\ket{{\boxed{{\color{blue}0}}}010}+\ket{{\boxed{{\color{blue}1}}}010}+\ket{{
\boxed{{\color{blue}0}}}011}+\ket{{\boxed{{\color{blue}1}}}011}\\
&+\ket{{\boxed{{\color{blue}0}}}100}+\ket{{\boxed{{\color{blue}1}}}100}+\ket{{
\boxed{{\color{blue}0}}}101}+\ket{{\boxed{{\color{blue}1}}}101}+\ket{{\boxed{{
\color{blue}0}}}110}+\ket{{\boxed{{\color{blue}1}}}110}+\ket{{
\boxed{{\color{blue}0}}}111}+\ket{{\boxed{{\color{blue}1}}}111})\ket{{
\boxed{{\color{blue}1}}}}
 \end{split}
 \label{eq:end_exam1}
\end{equation}
\end{subequations}\medskip

A measure on the flag qubit tells us that there is a value (here is $2$) which
gives $f(x)=1$.
\end{example}

It appears that we need to apply $n$ times the \texttt{NLE} gate. So, if we know
the values of $t$ qubits of our register (i.e., $t$ qubits have been measured or
are already disentangled to others, or the oracle acts on a subspace of $(n-t)$
qubits), the \texttt{NLE} gate will be repeated $(n-t)$ times. Let us see it
with another example and using the same conditions.
\begin{example}\label{exam2}
\begin{subequations}
 The value of the most significant qubit (MSQ) is known and it is $\ket{0}$ (a 
measure was done on it or the Hadamard gate was applied on it). Our system 
collapses to
\begin{equation}
\label{eq:MSQ1}
 \begin{split}
 \ket{\psi}&=\frac{1}{2\sqrt{2}}(\ket{0000}+\ket{0001}+\ket{0010}+\ket{0011}
+\ket {0100}+\ket{0101}+\ket{0110}+\ket{0111})\ket{0}\\
&=\ket{0}\left[\frac{1}{2\sqrt{2}}(\ket{000}+\ket{001}+\ket{010}+\ket{011}+\ket{
100}+\ket{101}+\ket{110}+\ket{111})\right]\ket{0}.
 \end{split}
\end{equation}
Else the oracle acts on the 3-qubit and gives $1$ for $\ket{010}$ which is a 
part of the values $2$ and $10$ in their binary forms ($0010$ and $1010$). The
system must be viewed as
\begin{equation}
 \ket{\psi}=\left(\frac{1}{\sqrt{2}}\ket{0}+\frac{1}{\sqrt{2}}\ket{1}\right)
\left[ \frac{1}{2\sqrt{2}}(\ket{000}+\ket{001}+\ket{010}+\ket{011}+\ket{100}
+\ket{101}+\ket{110}+\ket{111})\right] \ket{0}.\label{eq:MSQ2}
\end{equation}
If the MSQ, the fourth qubit, is noted $a$ ($\ket{a}=\ket{0}$
in Eq. (\ref{eq:MSQ1}) and $\ket{a}=\frac{1}{\sqrt{2}}\ket{0}+\frac{1}{\sqrt{2}}
\ket{1}$ in Eq. (\ref{eq:MSQ2})), the application of oracle operator gives
\begin{equation}
\ket{\Psi}=\frac{1}{2\sqrt{2}}\ket{a}(\ket{000}\ket{0}+\ket{001}\ket{0}
+\ket{010}\ket{0}+\ket{011}\ket{1}+\ket{100}\ket{0}+\ket{101}\ket{0}\\+\ket{110}
\ket{0}+\ket{111}\ket{0}).
\end{equation}
Next, proceeding like in example \ref{exam1} without as much as details, each
application of the \texttt{NLE} gate on the system gives
 \begin{equation}
  \begin{split}
   \ket{\Psi_1}=\frac{1}{2\sqrt{2}}\ket{a}&(\ket{010}\ket{1}+\ket{011}\ket{1}\\
&+\ket{000}\ket{0}+\ket{001}\ket{0}\\ &+\ket{100}\ket{0}+\ket{101}\ket{0}\\
&+\ket{110}\ket{0}+\ket{111}\ket{0}),
  \end{split}
 \end{equation}
 \begin{equation}
  \begin{split}
   \ket{\Psi_2}=\frac{1}{2\sqrt{2}}\ket{a}&(\ket{001}\ket{1}+\ket{011}\ket{1}\\
&+\ket{000}\ket{1}+\ket{010}\ket{1}\\&+\ket{100}\ket{0}+\ket{101}\ket{0}\\
&+\ket{110}\ket{0}+\ket{111}\ket{0}),
  \end{split}
 \end{equation}
 \begin{equation}
  \begin{split}
   \ket{\Psi_3}=\frac{1}{2\sqrt{2}}\ket{a}&(\ket{111}\ket{1}+\ket{011}\ket{1}\\
&+\ket{100}\ket{1}+\ket{000}\ket{1}\\&+\ket{101}\ket{1}+\ket{001}\ket{1}\\
&+\ket{110}\ket{1}+\ket{010}\ket{1})\\
 =\frac{1}{2\sqrt{2}}\ket{a}&(\ket{000}+\ket{001}+\ket{010}+\ket{011}
+\ket{100}+\ket{101}+\ket{110}+\ket{111})\ket{1}.
  \end{split}
 \end{equation}
 A measure on flag qubit gives the sought information.
\end{subequations}
\end{example}

\section{Concise algorithm for quantum associative memories}
\label{sec:Cal}

\subsection{Principles of algorithm}
We briefly describe here all the process of the QAM-NLSA. Like in the Rigui 
\etal paper's \cite{zhou2012}, the process of learning or storing patterns of 
our memory is done using an operator named $\mathtt{BDD}$ which is obtained 
while using the Binary Superposed Quantum Decision Diagram (BSQDD) proposed by 
Rosenbaum \cite{Rosenbaum2010}. BSQDD is computed by using any basis states 
$\ket{z}$ of Hilbert space of $2^n$ dimensions (not only $\ket{00\hdots0}$). 
According to Rosenbaum, the idea behind BSQDD is to represent a quantum 
superposition as a decision diagram where each node corresponds to a gate. The 
gate that corresponds to the node on each branch of the BSQDD is controlled by 
the path that was used to reach it from the root of the decision diagram. 
Thereby three steps are needed to construct a BSQDD. 
\begin{enumerate}
 \item \emph{Finding the unsimplified BSQDD by using Hadamard gates, Feymann 
gates and inverters} (see Fig. \ref{fig:bsqdd1} for a case of a register with 
fourth qubits). The number of node of this unsimplified BSQDD represents the 
upper bound on the number of the gates that will be needed to construct the 
quantum array generated by the BSQDD.
 \begin{figure}[!htbp]
 \centering
 \includegraphics[scale=0.5]{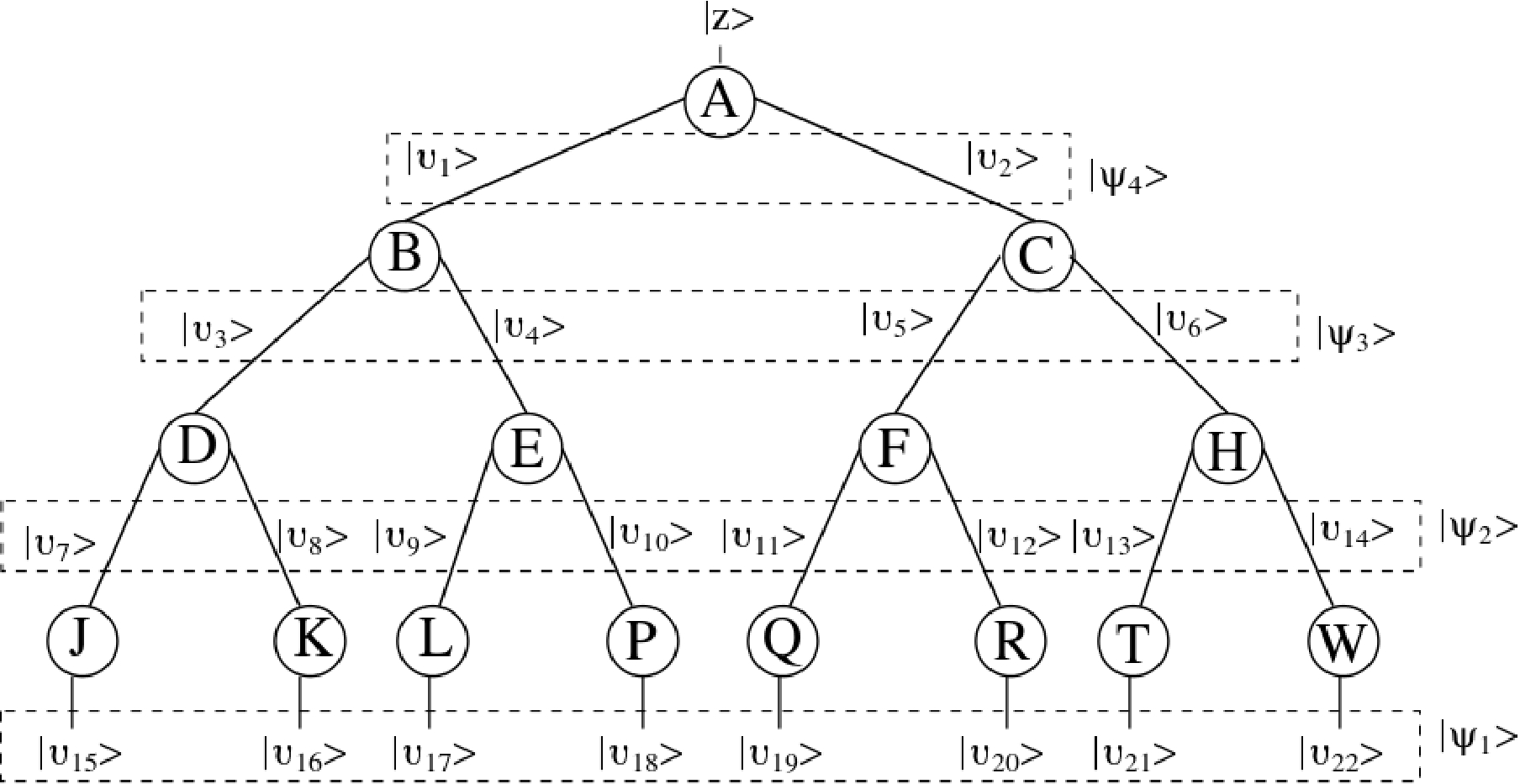}
  \caption{Unsimplified BSQDD. Each layer corresponds to a set of gates which 
act on a specific qubit. For example, the gate $A$ acts on the most significant 
qubit while gates $J$ to $W$ act on the least significant qubit. The state 
$\ket{\upsilon_i}$ is the non normalized state obtained when a specific gate 
acts on a specific qubit. The state $\ket{\psi_i}$, as the sum of non normalized 
states, is the normalized state obtained after the gates of the layer $i$ act on 
qubit $i$. The state $\ket{z}$ is the basis state used for starting and state 
$\ket{\psi_1}$ is the desired quantum supposed state.}
  \label{fig:bsqdd1}
 \end{figure}
 
\item \emph{Reducing the BSQDD to obtain the final BSQDD}. The goal is to have 
the lower bound on the number of quantum gates. To attempt this goal one needs 
to merge some nodes (gates) according to the links that can occur between 
qubits (like control qubit and target qubit). Fig. \ref{fig:merging} shows 
BSQDDs with merged nodes (Fig. \ref{fig:bsqdd2} and Fig. \ref{fig:bsqdd3}) and 
the final BSQDD (Fig. \ref{fig:bsqdd4}). The three BSQDDs in Fig. 
\ref{fig:merging} are equivalent.

\begin{figure}[!htbp]
\centering
 \leavevmode
 \subfloat[First merging nodes]{
 \includegraphics[scale=0.45]{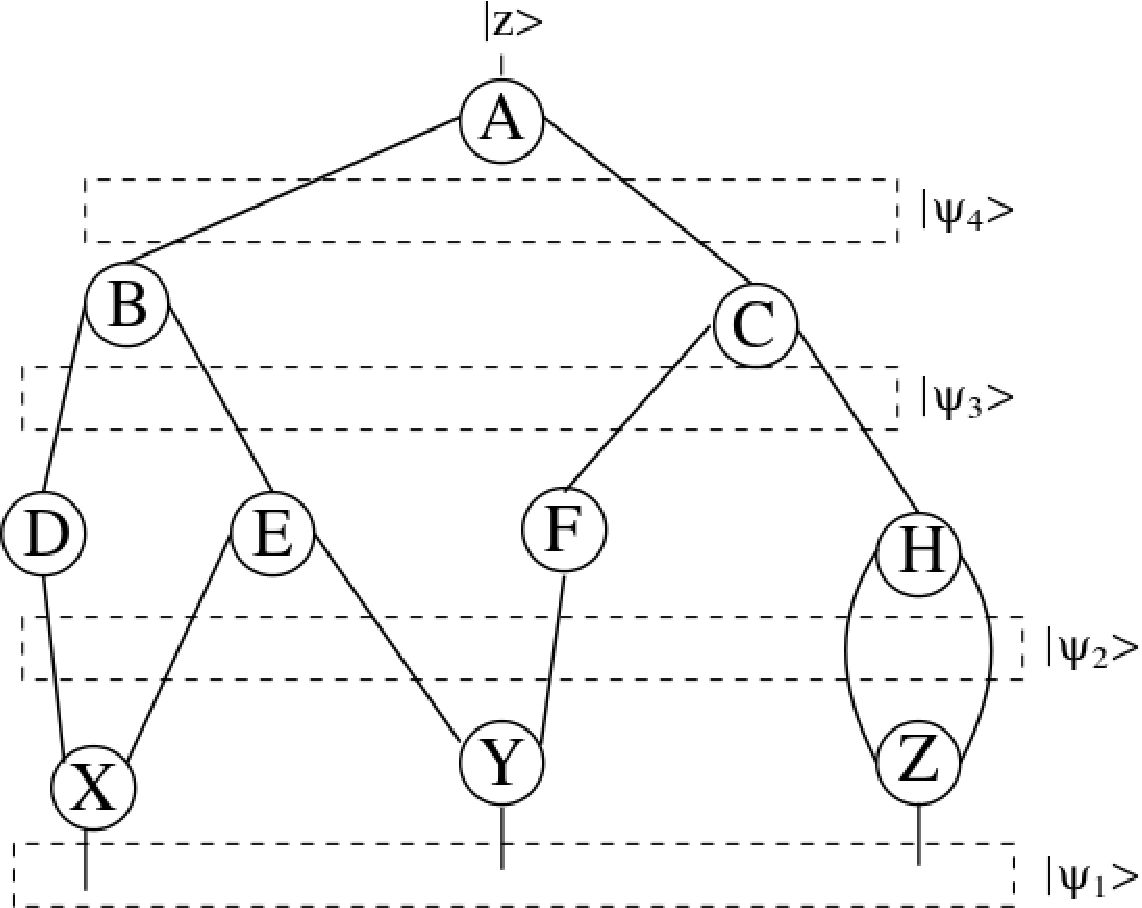}\label{fig:bsqdd2}}
 \subfloat[Second merging nodes]{
 \includegraphics[scale=0.45]{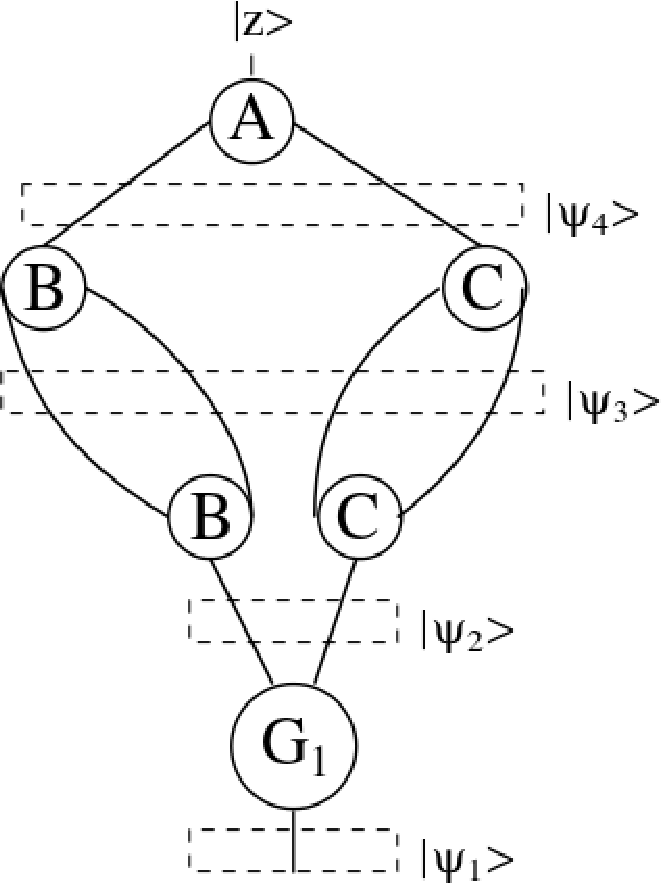}\label{fig:bsqdd3}}
 \subfloat[Final BSQDD]{
 \includegraphics[scale=0.45]{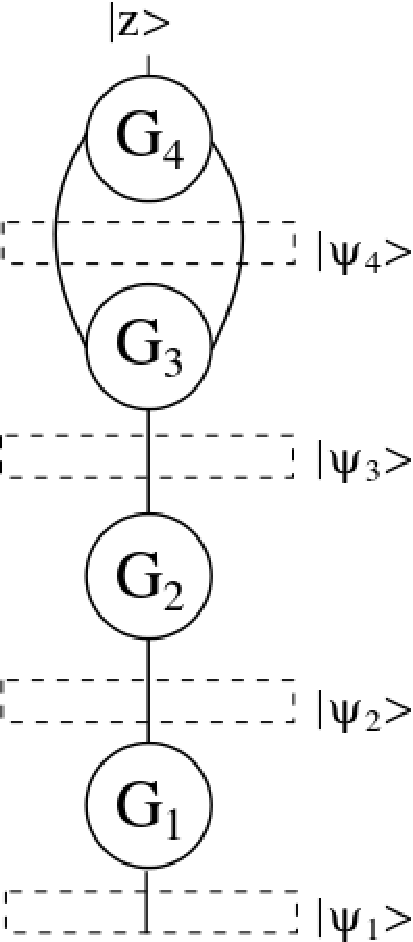}\label{fig:bsqdd4}}
 \caption{The merging nodes to obtain the final BSQDD by using the following 
two rules: in two different branches of different nodes which correspond to the 
same next same node, the nodes merge; in different branches of different nodes 
which generate the same branch, the branches merge.}
 \label{fig:merging}
 \end{figure}
 \item  \emph{Converting the BSQDD to a quantum array which generates the 
desired quantum state} (see Fig. \ref{fig:Quarray}).
\begin{figure}[!htbp]
 \[ \Qcircuit @C=1.5em @R=.9em{ 
&\ustick{\ket{z}}\gategroup{1}{2}{6}{2}{.0em}{--}&&
\ustick{\ket{\psi_4}}\gategroup{1}{4}{6}{4}{.0em}{--}&&\ustick{\ket{\psi_3}}
\gategroup{1}{6}{6}{6}{.0em}{--}&&\ustick{\ket{\psi_2}}
\gategroup{1}{8}{6}{8}{.0em}{--}&&\ustick{\ket{\psi_1}}
\gategroup{1}{10}{6}{10}{.0em}{--}&\\
 &\lstick{\ket{j_4}}&\gate{G_4}&\qw&\qw&\qw&\qw&\qw&\qw&\qw&\qw\\ 
&\lstick{\ket{j_3}}&\qw&\qw&\gate{G_3}&\qw&\qw&\qw&\qw&\qw&\qw\\
 &\lstick{\ket{j_2}}&\qw&\qw&\qw&\qw&\gate{G_2}&\qw&\qw&\qw&\qw\\
 &\lstick{\ket{j_1}}&\qw&\qw&\qw&\qw&\qw&\qw&\gate{G_1}&\qw&\qw\\
 &&&&&&&&&&\\
}
 \]
 \caption{The quantum array generated by the final BSQDD. The array is obtained 
by adding the gates for the nodes in each layer of the final BSQDD. The 
starting point is the last layer and new gates are always placed to the right 
of gates that have already been placed in quantum array. Therefore the first 
gate is $G_4$, while the last is $G_1$.}
 \label{fig:Quarray}
\end{figure}
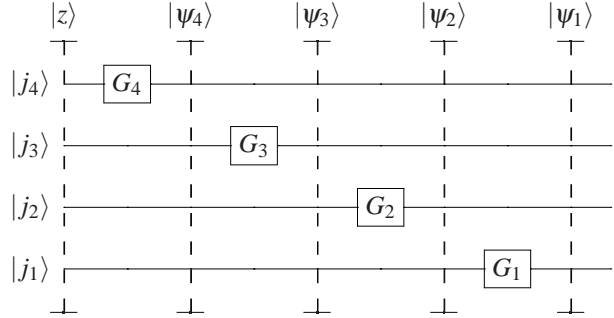
\end{enumerate}

\begin{figure}[!htbp]
\centering
\leavevmode
 \subfloat[Unsimplified BSQDD]{
 \includegraphics[scale=0.3]{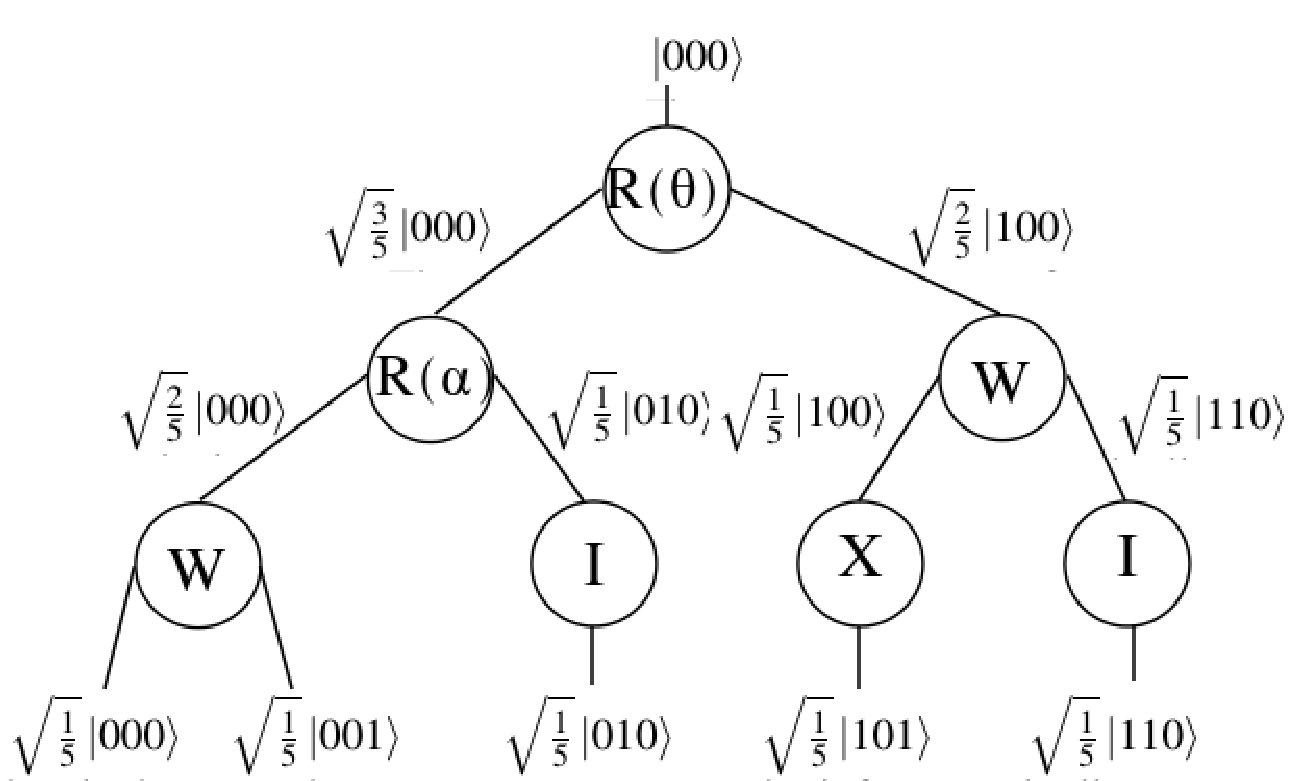}\label{fig:bsqdd5}}
 \subfloat[Final BSQDD]{
 \includegraphics[scale=0.5]{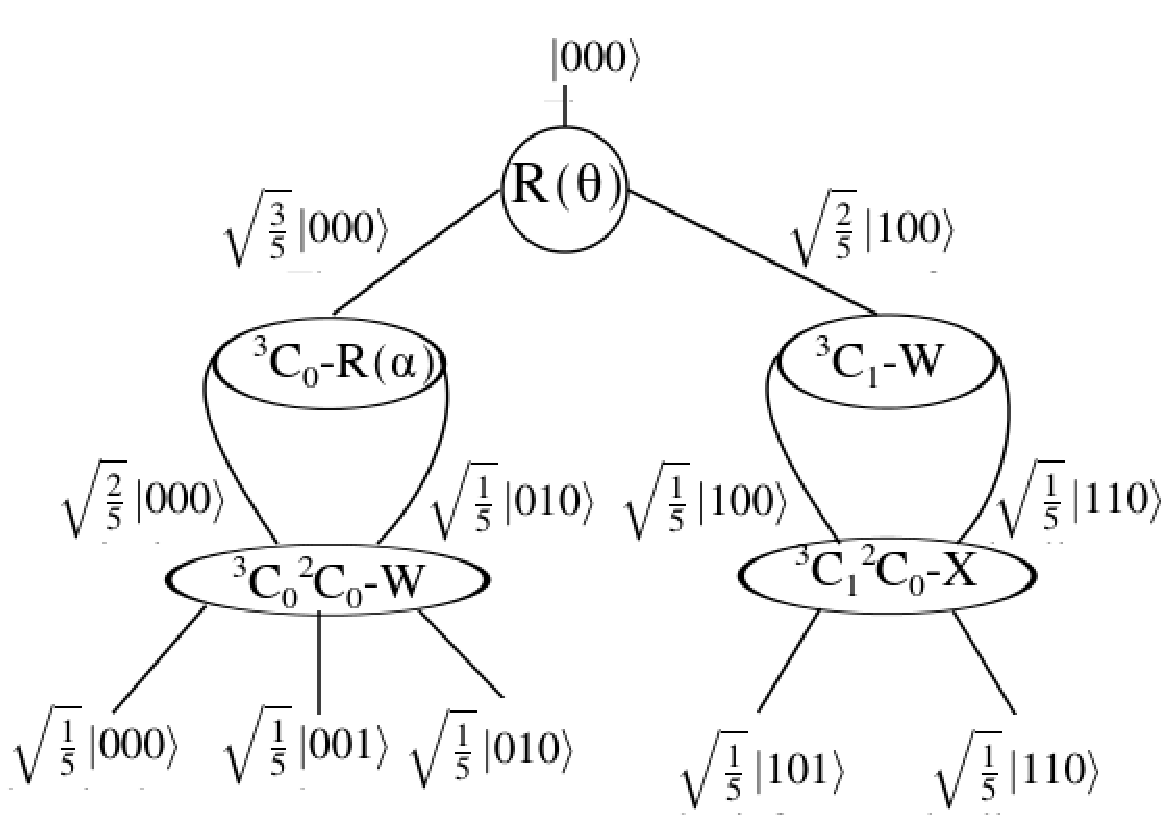}\label{fig:bsqdd6}}
 \\\subfloat[Corresponding quantum array]{
 $\Qcircuit @C=1.em @R=.7em{ 
&&\ustick{\ket{z}}
\gategroup{2}{4}{4}{4}{1.em}{--}\gategroup{1}{3}{5}{3}{.0em}{--}&&
\ustick{\ket{\psi_3}}\gategroup{1}{5}{5}{5}{.0em 
}{--}&&&&\ustick{\ket{\psi_2}}\gategroup{1}{9}{5}{9}{.0em 
}{--}\gategroup{2}{7}{4}{8}{2.4em}{--}&&&&\ustick{\ket{\psi_1}}
\gategroup{1}{13}{5}{13}{.0em}{--}\gategroup{2}{11}{4}{12}{1.em}{--}\\ 
&\lstick{\ket{0}}&\qw&\gate{\mathtt{R}(\theta)}&\qw&\qw&\ctrlo{1}&\ctrl{1}
&\qw&\qw&\ctrlo{1}&\ctrl{1}&\qw\\ 
&\lstick{\ket{0}}&\qw&\qw&\qw&\qw&\gate{\mathtt{R}(\alpha)}&\gate{\mathtt{W}} 
&\qw&\qw&\ctrlo{1}&\ctrlo{1}&\qw\\ 
&\lstick{\ket{0}}&\qw&\qw&\qw&\qw&\qw&\qw&\qw&\qw&\gate{\mathtt{W}}
&\gate{\mathtt{X}}&\qw\\
 &&& G_3&&&G_2&&&&G_1&&
 } $
 }
 \caption{BSQDD to obtain state 
$\ket{\psi_1}=\sqrt{\frac{1}{5}}(\ket{000}+\ket{010}+\ket{110}+\ket{001}+ 
\ket{101})$.}
 \label{fig:learn1}
 \end{figure}

\begin{example}
Fig. \ref{fig:learn1} gives the three steps which allow to construct the state 
$\sqrt{\frac{1}{5}}(\ket{000}+\ket{010}+\ket{110}+\ket{001}+\ket{101})$. The 
elementary gates used are respectively $\mathtt{R}(\theta) 
=\begin{pmatrix}\sqrt{\frac{3}{5}} & \sqrt{\frac{2}{5}}\\\sqrt{\frac{2}{5}} & 
-\sqrt{\frac{3}{5}}\end{pmatrix}$,  
$\mathtt{R}(\alpha)=\begin{pmatrix}\sqrt{\frac{2}{3}} & \frac{1}{\sqrt{3}}\\ 
\frac{1}{\sqrt{3}} & -\sqrt{\frac{2}{3}} \end{pmatrix}$, the Hadamard gate 
$\mathtt{W}$ and the NOT gate $\mathtt{X}$. Then 
$\ket{\psi_3}=\sqrt{\frac{3}{5}}\ket{000}+\sqrt{\frac{2}{5}}\ket{100}$ and 
$\ket{\psi_2}=\sqrt{\frac{2}{5}}\ket{000}+\frac{1}{\sqrt{5}}\ket{010}+\frac{1} 
{\sqrt{5}}\ket{100}+\frac{1}{\sqrt{5}}\ket{110}$.
\end{example}

\begin{figure}[!htbp]
\centering
\leavevmode
 \subfloat[Unsimplified BSQDD]{
 \includegraphics[scale=0.4]{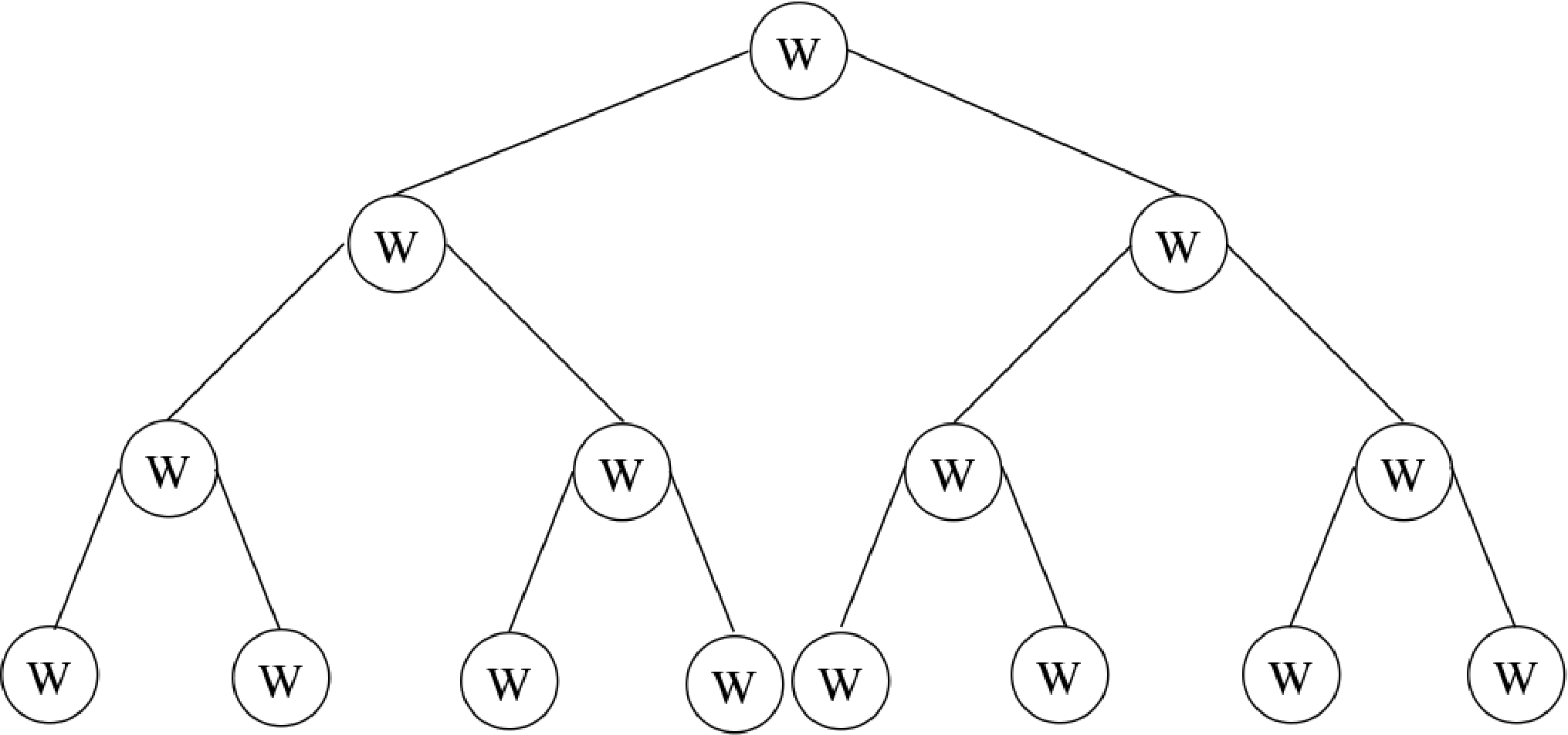}\label{fig:bsqdd7}}
 \subfloat[Final BSQDD]{
 \includegraphics[scale=0.4]{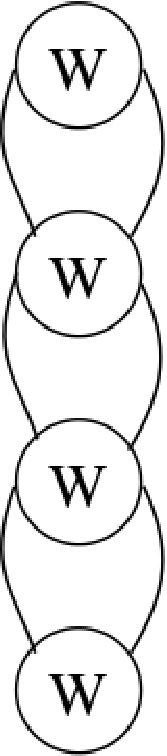}\label{fig:bsqdd8}}
 \caption{Two first steps needed to compute state used in example \ref{exam1}.}
 \label{fig:learn2}
 \end{figure}

Fig. \ref{fig:learn2} presents the two first steps needed to compute the state 
used in example \ref{exam1}. The corresponding quantum array is the first dashed 
box of Fig. \ref{fig:fig_algo2}.
 
The process of retrieving data is done by the quantum NLSA which allows us to 
have the information we want after a measure on the flag qubit not on the first 
register. However, it can be useful to measure the register, especially in case 
of multi-values which satisfy $f(x)=1$. But, as it appears in the previous 
section, we will get each $2^n$ values with the same probability. In the method 
proposed by Rigui \etal \cite{zhou2012} there are some ambiguities on how the 
system evolves and it is not clear on how a measure will give one of the sought 
patterns after the retrieving process.

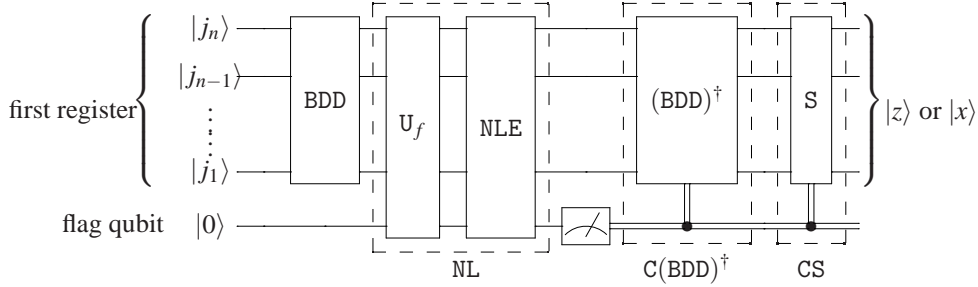
\begin{figure}[htbp]
\[ \Qcircuit @C=1.0em @R=.9em {
&&\ket{j_n}&&\qw&\multigate{4}{\mathtt{BDD}}&\multigate{5}{\mathtt{U}_f}
&\multigate{5}{\mathtt{NLE}}&\qw&\multigate{4}{(\mathtt{BDD})^{\dag}}&\qw&
\multigate{4}{\mathtt{S}}&\qw&\\&&\ket{j_{n-1}}&&\qw&\ghost{\mathtt{BDD}}
&\ghost{\mathtt{U}_f}&\ghost{\mathtt{NLE}}&\qw&\ghost{(\mathtt{BDD})^{\dag}}&\qw
&\ghost{\mathtt{S}}&\qw&\\&\lstick{\text{first register }\hspace{1.em}}
\gategroup{1}{1}{5}{1}{1em}{\{}&\vdots&& & & & & & &\rstick{\hspace{4em}
\ket{z}\text{ or }\ket{x}}\gategroup{1}{13}{5}{13}{1em}{\}}\\ &&\vdots& &  & & &
& &\\&&\ket{j_1}&&\qw&\ghost{\mathtt{BDD}}&\ghost{\mathtt{U}_f}
&\ghost{\mathtt{NLE}}&\qw&\ghost{(\mathtt{BDD})^{\dag}}&\qw&\ghost{\mathtt{S}}
&\qw&\\
&\lstick{\text{flag qubit }}&\ket{0}&&\qw&\qw&\ghost{\mathtt{U}_f}
&\ghost{\mathtt{NLE}}\gategroup{1}{7}{6}{8}{1em}{--}
&\meter&\control\cw\cwx\gategroup{1}{10}{6}{10}{1em}{--}&\cw&\control\cw\cwx
\gategroup{1}{12}{6}{12}{1em}{--}&\cw
\\
 &&&&&&\hspace{4.em}\mathtt{NL}&&&\mathtt{C}(\mathtt{BDD})^{\dag}&&\mathtt{CS}
&&
}
\]
 \caption{Schematic structure of QAM-NLSA. The $\mathtt{BDD}$ computes the 
learning process. The retrieving process is made by the gate $\mathtt{U}_f$ 
which marks the sought states, the gate $\mathtt{NLE}$ repeatedly computing the 
nonlinear evolution, the conditional gate $\mathtt{C}(\mathtt{BDD})^{\dag}$ 
bringing back the first register to its initial state $\ket{z}$ and the 
conditional operator $\mathtt{CS}$ which maps the first register to the sought 
state $\ket{x}$.}
 \label{fig:fig_algo}
\end{figure}

The figure \ref{fig:fig_algo} summarizes our QAM-NLSA where it is possible to 
retrieve one of the sought states in multi-values retrieving scheme when a 
measure on the first register is done.

\begin{enumerate}
\item The \textbf{learning process} is made by the operator $\mathtt{BDD}$.

\item The \textbf{retrieving process} is made by:
\begin{enumerate}
\item The operator $\mathtt{NL}$ which marks the sought states with
$\mathtt{U}_f$, computes repeatedly the nonlinear evolution $\mathtt{NLE}$ on
the system and disentangles the first register from the flag qubit.

\item The conditional operator $\mathtt{C}(\mathtt{BDD})^{\dag}$ which acts on
the first register and brings it back to its initial state $\ket{z}$ when the 
flag qubit is $\ket{1}$.

\item The operator
\begin{equation}
\mathtt{CS}=\mathbb{I}_{2^n}\otimes\ket{0}\bra{0}+\left(\mathbb{I}_{2^n}-(\ket{z
}\bra{z}+\ket{x}\bra{x})+\ket{x}
\bra{z}+\ket{z}\bra{x}\right)\otimes
\ket{1}\bra{1},
\label{eq:FtotalOp}
\end{equation}
which is a $(2^{n+1})\times(2^{n+1})$ conditional operator which maps the first
register to the sought state $\ket{x}$ when the flag qubit is $\ket{1}$. Put 
differently,
\begin{itemize}
 \item if the flag qubit is $\ket{0}$ nothing is done;
 \item if the flag qubit is $\ket{1}$ the $2^n\times2^n$ operator
 \begin{equation}
\mathtt{S}=\mathbb{I}_{2^n}-(\ket{z}\bra{z}+\ket{x}\bra{x})+\ket{x}
\bra{z}+\ket{z}\bra{x}
  \label{eq:partOp}
 \end{equation}
 is applied on the first register.
\end{itemize}

It is noteworthy that in the case of multi-patterns retrieving the sought state
$\ket{x}$ can be a supposed state of all the sought states (for example
$\ket{x}=\frac{1}{\sqrt{2}}(\ket{0010}+\ket{1010})$ in the example \ref{exam2}).

\item The system is observed by making a measurement on the first register
and/or on the flag qubit to erase any ambiguity.
\end{enumerate}
\end{enumerate}
We also point out the fact that as the BSQDD method can compute any sought
state, it can be useful to compute the operator $\mathtt{CS}$. Indeed, in the
case of a complex sought state $\ket{x}$ where the Hadamard gates or other 
methods are inadequate, the BSQDD method can allow us to have the appropriate 
form of the operator $\mathtt{S}$.

\begin{example}
 The example \ref{exam1} suggests the operator $\mathtt{S}=\mathbb{I}_2
\otimes \mathbb{I}_2\otimes\mathtt{X}\otimes \mathbb{I}_2$. Therefore, the
$\mathtt{CX}$ gate acts only on the second qubit, while the other qubits are 
unchanged. The figure \ref{fig:fig_algo2} gives the evolution of the system.
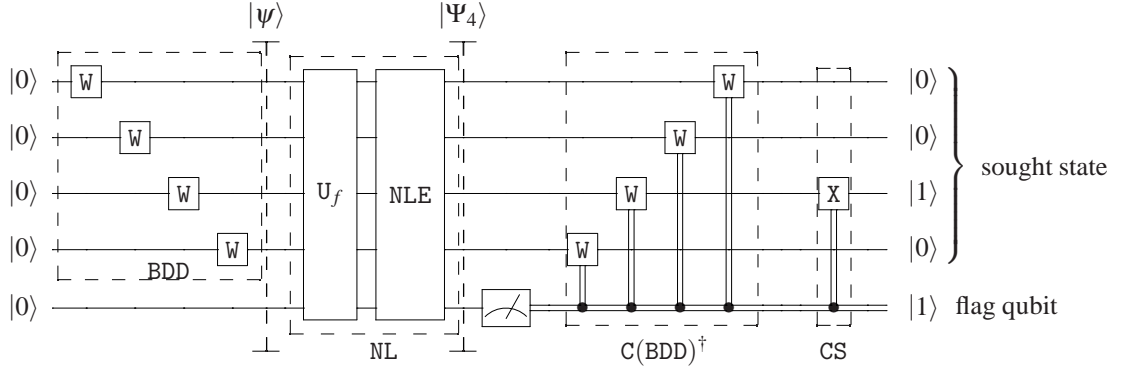
\begin{figure}[htbp]
\[ \Qcircuit @C=.7em @R=.9em { 
&&&&&&\ustick{\ket{\psi}}&&& &\ustick{\ket{\Psi_4}}&&&&&&&&&\\
&\lstick{\ket{0}}&\gate{\mathtt{W}}&\qw&\qw&\qw&\qw&\qw&\multigate{4}{\mathtt{U}
_f}
&\multigate{4}{\mathtt{NLE}}&\qw&\qw&\qw&\qw&\qw&\qw&\gate{\mathtt{W}}
&\qw&\qw&\qw&\qw&\qw
&\qw & &\ket{0}&\\
 &\lstick{\ket{0}}&\qw&\gate{\mathtt{W}}&\qw&\qw&\qw&\qw&\ghost{\mathtt{U}_f}&
 \ghost{\mathtt{NLE}}&\qw&\qw&\qw&\qw&\qw&\gate{\mathtt{W}}&\qw&\qw&\qw&\qw&\qw
&\qw&\qw&
&\ket{0}&\\
 &\lstick{\ket{0}}&\qw&\qw&\gate{\mathtt{W}}&\qw&\qw&\qw&\ghost{\mathtt{U}_f}& 
\ghost{\mathtt{NLE}}&\qw&\qw&\qw&\qw&\gate{\mathtt{W}}&\qw&\qw&\qw&\qw&\qw&\gate
{\mathtt{X}}&\qw&\qw&&\ket{1}&&\ustick{\hspace*{6em}\text{ sought state}} 
\gategroup{2}{26}{5}{26}{1em}{\}}\\
&\lstick{\ket{0}}&\qw&\dstick{\hspace{2.5em}
\mathtt{BDD}}
\qw&\qw&\gate{ \mathtt{W}}&\qw&
 \qw&\ghost{\mathtt{U}_f}&\ghost{\mathtt{NLE}}&\qw&\qw&\qw&\gate{\mathtt{W}}
&\qw&\qw&\qw&\qw
&\qw& \qw&\qw&\qw&\qw&&\ket{0}&\\ 
&\lstick{\ket{0}}&\qw\gategroup{2}{3}{5}{6}{1em}{--}
&\qw\gategroup{1}{7}{7}{7}{0em}{--}&\qw&\qw&\qw&\qw&\ghost{ 
\mathtt{U}_f}&\ghost{\mathtt{NLE}}&\qw\gategroup{1}{11}{7}{11}{0em}{--}
&\meter&\cw\gategroup{2}{9}{6}{10}{1em}{--}
&\control\cw\cwx&\control\cw\cwx[-2]&\control\cw\cwx[-3]
&\control\cw\cwx[-4]&\cw&\cw&\cw\gategroup{2}{14}{6}
{17}{1em}{--}&\control\cw\cwx[-2] \gategroup{2}{21}{6}{21}{1em}{--}
&\cw&\cw&&\ket{1}&\rstick{\text{flag qubit}}\\
&&&&&&&&\hspace{4.em}\mathtt{NL}&&&&&&\hspace{2.5em}\mathtt{C}(\mathtt{BDD})^{
\dag} &&&&&\hspace{2.5em}\mathtt{CS}
 }
\]
 \caption{Schematic structure of QAM-NLSA of example \ref{exam1}. 
$\mathtt{W}$ is the Hadamard gate which puts (reverses to its initial state 
respectively) the first register in the superposed state before (after 
respectively) the $\mathtt{NL}$ computing the nonlinear algorithm acts. Note 
that states $\ket{\psi}$ and $\ket{\Psi_4}$ are those give by Eq. 
(\ref{eq:start_exam1}) and Eq. (\ref{eq:end_exam1}) respectively.}
 \label{fig:fig_algo2}
\end{figure}
\end{example}

\subsection{Analysis of the complexity of the nonlinear evolution algorithm}
\label{sec:pre}

All the above description was made with the assumption that there is at most
one value $x$ for which $f(x)=1$. Let us now consider the case where there can
be more than one value satisfying $f(x)=1$. In the simple case where there are
at most two values satisfying $f(x)=1$, the state (\ref{eq:equaO}) must be 
rewritten as
\begin{multline}
\ket{\Psi}=\frac{1}{\sqrt{2^n}}[\sum_{\substack{j_nj_{n-1}\hdots j_1=0\\
j_nj_{n-1}\hdots j_2\neq i_ni_{n-1}\hdots i_2\\j_nj_{n-1}\hdots j_2\neq
e_ne_{n-1}\hdots e_2}}^1\ket{j_nj_{n-1}\hdots j_1}
\ket{0}+\ket{i_ni_{n-1}\hdots(1-i_1)}\ket{0}+\ket{i_ni_{n-1}\hdots i_1}\ket{1}\\
+\ket{e_ne_{n-1}\hdots(1-e_1)}\ket{0}+\ket{e_ne_{n-1}\hdots e_1}\ket{1}].
\label{eq:equa2}
\end{multline}

Highlighting the LSQ of the first register and the flag qubit, the second part
of state (\ref{eq:equa2}) must be in one of the following states:
\begin{itemize}
\item if $i_ni_{n-1}\hdots i_2\neq e_ne_{n-1}\hdots e_2$,
\begin{subequations}
 \begin{align}
  &\ket{10}+\ket{01}+\ket{10}+\ket{01},\label{eq:NL1}\\
  &\ket{10}+\ket{01}+\ket{00}+\ket{11},\label{eq:NLO1}\\
  &\ket{00}+\ket{11}+\ket{10}+\ket{01},\label{eq:NLO2}\\
  &\ket{00}+\ket{11}+\ket{00}+\ket{11},\label{eq:NLT1}
 \end{align}
 \label{eq:NLe1}
\end{subequations}
\item   or if $i_ni_{n-1}\hdots i_2= e_ne_{n-1}\hdots e_2$,
 \begin{equation}
  \ket{01}+\ket{11},
  \label{eq:NL11}
 \end{equation}
according to the fact that there is no repetition of value.
\end{itemize}

The action of the \texttt{NLE} gate on states (\ref{eq:NLe1}) is the same as 
the one described in section \ref{sec:Snl}. Taking a careful look on states 
(\ref{eq:NLe1}), it seems that the \texttt{NLE} was already applied one time 
and 
this suggests that the \texttt{NLE} gate will be repeated $(n-1)$ times. The 
state (\ref{eq:NL11}) also supposes that the \texttt{NLE} gate was already 
applied thus the \texttt{NLE} gate will begin on the second LSQ and will be 
repeated $(n-1)$ times. Therefore, if there are at most two values $x$ for 
which 
$f(x)=1$, the number of steps of the QAM-NLSA is
\begin{subequations}
\begin{equation}
 \mathcal{O}(n-1),
\end{equation}
as the \texttt{NLE} gate  starts on the second LSQ. It is easy to find that in 
the case where there are at most three values $x$ satisfying $f(x)=1$, when we 
start the repeated action of the \texttt{NLE} on the second LSQ of the first 
register, the number of steps of the QAM-NLSA is also
\begin{equation}
 \mathcal{O}(n-1).
\end{equation}
\end{subequations}

According to the above observation, if there are at most $m$ values $x$ for 
which $f(x)=1$ and $r=\mathtt{int}(\log_2{m})$, that is the integer part of 
$\log_2{m}$, the \texttt{NLE} gate action starts on the $(r+1)^{th}$ LSQ of the 
first register and will be repeated $(n-r)$ times. Thus, the number of steps of 
the QAM-NLSA is
\begin{equation}
 \mathcal{O}(n-r).
 \label{eq:step}
\end{equation}

\begin{example}\label{exam3}
In order to enlighten the result (\ref{eq:step}), let us consider the 
parameters of example \ref{exam1}. The marked states are $\ket{2}=\ket{0010}$, 
$\ket{5}=
\ket{0101}$, $\ket{8}=\ket{1000}$, $\ket{10}=\ket{1010}$, $\ket{11}=\ket{1011}$,
$\ket{13}=\ket{1101}$ and $\ket{15}=\ket{1111}$. The number of marked states is
$m=7$ and $\log_2{m}=2.807$. Consequently, $r=2$; and the \texttt{NLE} gate 
action starts on the third LSQ of the register. Let us check that in detail. 
As 
in the example \ref{exam1}, we start with
\begin{subequations}
\begin{multline}
 \ket{\psi}=\frac{1}{4}(\ket{0000}+\ket{0001}+\ket{0010}+\ket{0011}+\ket{0100
}+\ket{0101}+\ket{0110}+\ket{0111}\\+\ket{1000}+\ket{1001}+\ket{1010}+\ket{1011}
+\ket{1100} +\ket{1101}+\ket{1110}+\ket{1111})\ket{0}.
\end{multline}
 The action of the oracle operator $\mathtt{U}_f$ yields
 \begin{multline}
 \ket{\Psi}=\frac{1}{4}(\ket{0000}\ket{0}+\ket{0001}\ket{0}+\ket{0010}\ket{
1}+\ket{0011}\ket{0}+\ket{0100}\ket{0}+\ket{0101}\ket{1}+\ket{0110}\ket{0}
\\+\ket{0111}\ket{0}+\ket{1000}\ket{1}+\ket{1001}\ket{0}+\ket{1010}\ket{
1}+\ket{1011}\ket{1}+\ket{1100}\ket{0}+\ket{1101}\ket{1}\\+\ket{1110}\ket{0}
+\ket{1111}\ket{1}).
\end{multline}
\end{subequations}
\begin{subequations}
 \begin{minipage}[c]{.48\linewidth}
Highlighting the third qubit,
\begin{equation}
\begin{split}
 \ket{\Psi}=\frac{1}{4}&(\ket{0{
\boxed{{\color{blue}0}}}00}\ket{{\boxed{{\color{blue}0}}}}+\ket{0{\boxed{{\color
{blue}1}}}00}\ket{{\boxed{{ \color{blue}0}} }}\\ &+\ket{0{
\boxed{{\color{blue}0}}} 01}\ket{{
\boxed{{\color{blue}0}}}}+\ket{0{\boxed{{\color{blue}1}}}01}\ket{{\boxed{{\color
{blue}1}}}}\\ &+\ket{0{ \boxed{{\color{blue}0}}} 10}\ket{{
\boxed{{\color{blue}1}}}}+\ket{0{\boxed{{\color{blue}1}}}10}\ket{{\boxed{{\color
{blue}0}}}}\\ &+\ket{0{ \boxed{{\color{blue}0}}} 11}\ket{{
\boxed{{\color{blue}0}}}}+\ket{0{\boxed{{\color{blue}1}}}11}\ket{{\boxed{{\color
{blue}0}}}}\\ &+\ket{1{
\boxed{{\color{blue}0}}}01}\ket{{\boxed{{\color{blue}0}}}}+\ket{1{\boxed{{\color
{blue}1}}}01}\ket{{\boxed{{\color{blue}1}}}} \\ &+\ket{1{
\boxed{{\color{blue}0}}}11}
\ket{{\boxed{{\color{blue}1}}}}+\ket{1{\boxed{{\color{blue}1}}}11}\ket{{\boxed{{
\color{blue}1}}}} \\ &+\ket{1{
\boxed{{\color{blue}0}}}00}\ket{{\boxed{{\color{blue}1}}}}+\ket{1{\boxed{{\color
{blue}1}}}00}\ket{{\boxed{{\color{blue}0}}}}\\ &+\ket{1{
\boxed{{\color{blue}0}}}10}\ket{{\boxed{{\color{blue}1}}}}+\ket{1{\boxed{{\color
{blue}1}}}10}\ket{{\boxed{{\color{blue}0}}}})
 \end{split}
\end{equation}
\end{minipage} \hfill\begin{minipage}[c]{.48\linewidth}
and applying the \texttt{NLE} gate the first time produces
\begin{equation}
\begin{split}
 \ket{\Psi_1}=\frac{1}{4}&(\ket{0{
\boxed{{\color{blue}0}}}00}\ket{{\boxed{{\color{blue}0}}}}+\ket{0{\boxed{{\color
{blue}1}}}00}\ket{{\boxed{{ \color{blue}0}} }}\\ &+\ket{0{
\boxed{{\color{blue}0}}} 01}\ket{{
\boxed{{\color{blue}1}}}}+\ket{0{\boxed{{\color{blue}1}}}01}\ket{{\boxed{{\color
{blue}1}}}}\\ &+\ket{0{ \boxed{{\color{blue}0}}} 10}\ket{{
\boxed{{\color{blue}1}}}}+\ket{0{\boxed{{\color{blue}1}}}10}\ket{{\boxed{{\color
{blue}1}}}}\\ &+\ket{0{ \boxed{{\color{blue}0}}} 11}\ket{{
\boxed{{\color{blue}0}}}}+\ket{0{\boxed{{\color{blue}1}}}11}\ket{{\boxed{{\color
{blue}0}}}}\\ &+\ket{1{
\boxed{{\color{blue}0}}}01}\ket{{\boxed{{\color{blue}1}}}}+\ket{1{\boxed{{\color
{blue}1}}}01}\ket{{\boxed{{\color{blue}1}}}} \\ &+\ket{1{
\boxed{{\color{blue}0}}}11}
\ket{{\boxed{{\color{blue}1}}}}+\ket{1{\boxed{{\color{blue}1}}}11}\ket{{\boxed{{
\color{blue}1}}}} \\ &+\ket{1{
\boxed{{\color{blue}0}}}00}\ket{{\boxed{{\color{blue}1}}}}+\ket{1{\boxed{{\color
{blue}1}}}00}\ket{{\boxed{{\color{blue}1}}}}\\ &+\ket{1{
\boxed{{\color{blue}0}}}10}\ket{{\boxed{{\color{blue}1}}}}+\ket{1{\boxed{{\color
{blue}1}}}10}\ket{{\boxed{{\color{blue}1}}}})
 \end{split}
\end{equation}
\end{minipage}\medskip

\begin{minipage}[c]{.48\linewidth}
Finally, when taking a look on the MSQ, this is what we have:
\begin{equation}
\begin{split}
 \ket{\Psi_1}=\frac{1}{4}&(\ket{{\boxed{{\color{blue}0}}}000}\ket{{\boxed{{
\color{blue}0}}}}+\ket{{\boxed{{\color{blue}1}}}000}\ket{{\boxed{{\color{blue}1}
}}}\\ &+\ket{{\boxed{{\color{blue}0}}}001}\ket{{\boxed{{\color{blue}1}}}}+\ket{{
\boxed{{\color{blue}1}}}001}\ket{{ \boxed{{\color {blue}0}}}}\\
&+\ket{{\boxed{{\color{blue}0}}} 010}\ket{{
\boxed{{\color{blue}1}}}}+\ket{{\boxed{{\color{blue}1}}}010}\ket{{\boxed{{\color
{blue}0}}}}\\
&+\ket{{\boxed{{\color{blue}0}}}011}\ket{{\boxed{{\color{blue}0}}}}+\ket{{
\boxed {{\color{blue}1}}}011}\ket{{ \boxed{{\color{blue}1}}}}\\
&+\ket{{\boxed{{\color{blue}0}}} 100}\ket{{
\boxed{{\color{blue}0}}}}+\ket{{\boxed{{\color{blue}1}}}100}\ket{{\boxed{{\color
{blue}1}}}}\\ &+\ket{{\boxed{{\color{blue}0}}} 101}\ket{{
\boxed{{\color{blue}1}}}}+\ket{{\boxed{{\color{blue}1}}}101}\ket{{\boxed{{\color
{blue}0}}}}\\ &+\ket{{\boxed{{\color{blue}0}}} 110}\ket{{
\boxed{{\color{blue}1}}}}+\ket{{\boxed{{\color{blue}1}}}110}\ket{{\boxed{{\color
{blue}0}}}}\\ &+\ket{{ \boxed{{\color{blue}0}}} 111}\ket{{
\boxed{{\color{blue}0}}}}+\ket{{\boxed{{\color{blue}1}}}111}\ket{{\boxed{{\color
{blue}1}}}})
\end{split}
\end{equation}
\end{minipage} \hfill\begin{minipage}[c]{.48\linewidth}
and applying the \texttt{NLE} gate the last time produces
\begin{equation}
\begin{split}
 \ket{\Psi_2}=\frac{1}{4}&(
 \ket{{\boxed{{\color{blue}0}}}000}+\ket{{\boxed{{\color{blue}1}}}000}\\
 &+\ket{{\boxed{{\color{blue}0}}}001}+\ket{{\boxed{{\color{blue}1}}}001}\\
 &+\ket{{\boxed{{\color{blue}0}}}010}+\ket{{\boxed{{\color{blue}1}}}010}\\
 &+\ket{{\boxed{{\color{blue}0}}}011}+\ket{{\boxed{{\color{blue}1}}}011}\\
 &+\ket{{\boxed{{\color{blue}0}}}100}+\ket{{\boxed{{\color{blue}1}}}100}\\
 &+\ket{{\boxed{{\color{blue}0}}}101}+\ket{{\boxed{{\color{blue}1}}}101}\\
 &+\ket{{\boxed{{\color{blue}0}}}110}+\ket{{\boxed{{\color{blue}1}}}110}\\
 &+\ket{{\boxed{{\color{blue}0}}}111}+\ket{{\boxed{{\color{blue}1}}}111})\ket{{
\boxed{{ \color{blue}1}}}}
 \end{split}
\end{equation}
\end{minipage}
\end{subequations}\medskip

It effectively appears that the \texttt{NLE} gate was repeated $(n-r)=4-2=2$
times.
\end{example}

If the values of $t$ qubits of our first register are known (i.e., $t$ qubits 
have been measured or are already disentangled to others, or the oracle acts on 
a subspace of $(n-t)$ qubits) and there is at most $m$ values $x$ for which 
$f(x)=1$, the \texttt{NLE} gate will act repeatedly $((n-t)-r)$ times that 
starts on the $(r+1)^{th}$ LSQ. As the $t$ qubits which are already known will 
be 
ignored, it is clear that $m\leq2^{n-t}$. Consequently, the number of steps of 
the QAM-NLSA is
\begin{equation}
 \mathcal{O}((n-t)-r).
 \label{eq:step1}
\end{equation}

Now, if in the first register which is an $n$-qubit system, the computed 
patterns are $p\leq2^n$ and we stated $b=\mathtt{ceil}(\log_2{p})$, i.e. the 
least integer greater or equal to $\log_2{p}$, $m$ the number of value $x$ for 
which $f(x)=1$. The \texttt{NLE} gate will act repeatedly $(b-r)$ times. 
Therefore, the number of steps of the QAM-NLSA is
\begin{equation}
 \mathcal{O}(b-r)
\end{equation}
for which the upper bound is the Eq.(\ref{eq:step}). Note that the starting 
point of the \texttt{NLE} gate action will always be the $(r+1)^{th}$ LSQ.

If we know the values of $t$ qubits of our first register, it supposes that
we must view the system in terms of $q\leq p\leq2^n$ patterns; consequently the
number of values $m$ for which $f(x)=1$ is $m\leq q$. For $c=\mathtt{ceil}
(\log_2{q})$, the \texttt{NLE} gate  will act repeatedly $(c-r)$ times.
Therefore, the number of steps of the QAM-NLSA takes the general form
\begin{equation}
 \mathcal{O}(c-r).
\end{equation}

\begin{example}
 \begin{itemize}
\item If we consider again the parameters of the example \ref{exam1}, we find
that $p=16$, the number of known qubits is $t=0$. Consequently $q=p=16$, and
$\log_2{q}=\log_2{16}=4.0$, thus $c=4$, $m=1$ and $r=0$. The \texttt{NLE} gate 
will act repeatedly, $(c-r)=4-0=4$ times.

\item In the example \ref{exam2}, $t=1$. Consequently $q=8$ according to the
assumption taken in this example. $\log_2{q}=\log_2{8}=3.0$, thus $c=3$.
$m=1$, then $r=0$. The \texttt{NLE} gate  will act repeatedly, $(c-r)=3-0=3$
times.

\item In the example \ref{exam3}, $t=0$ and $q=16$ but $m=7$. Then 
$\log_2{m}=2.807$. That is $r=2$. The \texttt{NLE} gate  will act repeatedly
$(c-r)=4-2=2$ times.
\end{itemize}

\end{example}

It is noteworthy that when the state $\ket{01}+\ket{11}$ appears the last time,
the \texttt{NLE} gate acts repeatedly. The state $\ket{01}+\ket{11}$ then does 
not evolve. Its nonlinear
evolution must be like that of the state (\ref{eq:NL}). Such a nonlinear 
evolution of the state $\ket{01}+\ket{11}$ is described as follows: 
\begin{itemize}
 \item[\textbf{Step \ref{alg:Balgo13}.}] Apply operator $\mathtt{U}$:
 \begin{equation}
\mathtt{U}(\ket{01}+\ket{11})=\frac{1}{\sqrt{2}}(\ket{00}+\ket{01}+\ket{10}-\ket
{11}).
 \end{equation}

\item[\textbf{Step \ref{alg:algo14}.\ref{alg:algo141}.}]  Apply the nonlinear
operator $\mathtt{NL}^-$:
  \begin{equation}
\mathtt{NL}^-[\frac{1}{\sqrt{2}}(\ket{00}+\ket{01}+\ket{10}-\ket{11})]
=\sqrt{2}\ket{0}(\delta\ket{0}+\epsilon\ket{1}),\label{eq:equan-01}
\end{equation}
where $\delta,\epsilon\in\mathbb{C},\,|\delta|^2+|\epsilon|^2=1$. As we see on
the state (\ref{eq:equan-01}), the action of the nonlinear operator
$\mathtt{NL}^-$ is also not specified like on the state (\ref{eq:equan-0}).

\item[\textbf{Step \ref{alg:algo14}.\ref{alg:algo142}.}] Apply the second
nonlinear operator $\mathtt{NL}^+$:
\begin{equation}
\mathtt{NL}^+[\sqrt{2}\ket{0}(\delta\ket{0}+\epsilon\ket{1})]=\sqrt{2}\ket{00}.
\label{eq:equaT1}
\end{equation}
The general form of the unitary matrix $\mathtt{NL}^+$ which maps the generic 
1-qubit $\delta\ket{0}+\epsilon\ket{1}$ to $\ket{0}$ is
\begin{equation}
 \mathtt{\Pi}=\begin{pmatrix}
    \delta^{\ast} & \epsilon^{\ast}\\
    \mp\gamma\epsilon & \pm\gamma\delta
   \end{pmatrix},\,\,\gamma=\pm i\text{ or }\pm 1\;(i^2=-1),\label{eq:matri1}
\end{equation}
where $\,\delta,\,\epsilon\in\mathbb{C},\,|\delta|^2+|\epsilon|^2=1$.

\item[\textbf{Step \ref{alg:Ealgo13}.}] Apply the NOT gate $\mathtt{X}$
on the flag qubit and the Hadamard gate $\mathtt{W}$ on the first
qubit.
\end{itemize}
We summarize below this nonlinear evolution of state $\ket{01}+\ket{11}$ and 
give 
its corresponding circuit (see Fig. \ref{fig:fig6})
 \begin{equation}
\ket{01}+\ket{11}\xrightarrow{\mathtt{U}}\frac{1}{\sqrt{2}}(\ket{00}+\ket{01}
+\ket{10} -\ket{11})\xrightarrow{\mathtt{NL}^-}\sqrt{2}
\ket{0}(\delta\ket{0}+\epsilon\ket{1})\xrightarrow{\mathtt{NL}^+}\sqrt{2}\ket{00
}\xrightarrow{\mathtt{W}\otimes\mathtt{X}}\ket{01}+\ket{11}.
\label{eq:e11}
 \end{equation}
 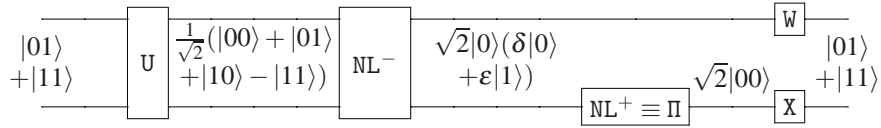
\begin{figure}[H]
\[  \Qcircuit @C=1.6em @R=2.em{
 &\dstick{\begin{matrix}\ket{01}\\+\ket{11}\end{matrix}}&\qw&\multigate{1}{
\mathtt{U}} &\qw&\qw&\qw&\multigate{1}{\mathtt{NL}^-}&\qw&\qw&\qw&\qw&\qw
&\gate{\mathtt{W}}&\dstick{\begin{matrix}\ket{01}\\+\ket{11}\end{matrix}}\qw\\
&&\qw&\ghost{\mathtt{U}}&\qw&\ustick{\begin{matrix}\frac{1}{\sqrt{2}}(\ket{00}
+\ket{01}\\+\ket{10}-\ket{11})\end{matrix}}\qw&\qw&\ghost{\mathtt{NL}^-}&\qw&
\ustick{\begin{matrix}\sqrt{2}
\ket{0}(\delta\ket{0}\\+\epsilon\ket{1})\end{matrix}}\qw&\qw&\gate{\mathtt{NL}^+
\equiv\mathtt{\Pi}}&\ustick{\sqrt{2}\ket{00}}\qw&\gate{\mathtt{X}}&\qw
  }
\]
  \caption{Equivalent circuit of the nonlinear evolution (\ref{eq:e11}).}
  \label{fig:fig6}
 \end{figure}

\begin{algorithm}
	\caption{Algorithm of QAM-NLSA}
	\begin{algorithmic}[1]
		\STATE Store patterns using the operator $\mathtt{BDD}$ and put 
the flag
		qubit to $\ket{0}$
		\STATE Apply the oracle $\mathtt{U}_f$
		\STATE \textbf{repeat} $(c-r)$ times  step (\ref{alg:begin}) to 
step
		(\ref{alg:end}) (i.e., one time per qubit of the first register 
starting from
		$(r+1)^{th}$ qubit with the flag qubit)
		\STATE\hspace{1em}\label{alg:begin} Apply the unitary operator 
$\mathtt{U}$
		\STATE \begin{enumerate}
			\item Apply the nonlinear operator $\mathtt{NL}^-$
			\item Apply the nonlinear operator $\mathtt{NL}^+$
		\end{enumerate}
		\STATE\hspace{1em}\label{alg:end} Apply the Hadamard operator 
$\mathtt{W}$ on
		the qubit of the first register and the NOT operator 
$\mathtt{X}$ on the flag
		qubit
		\STATE Observe the flag qubit
		\STATE \textbf{if} the flag qubit is in state $\ket{0}$
		\STATE\hspace{1em} Conclude
		\STATE \textbf{else}\begin{enumerate}
			\item Apply the conditional operator 
$\mathtt{C}(\mathtt{BDD})^{\dag}$ to kick
			back the first register to its initial state
			\item Apply the conditional operator $\mathtt{CS}$ to 
flip the register to 
			the sought state
			\item Observe the first register
		\end{enumerate}
	\end{algorithmic}
	\label{alg:algo2}
\end{algorithm}

Finally, the Algorithm (\ref{alg:algo2}) can describe the QAM-NLSA where
\begin{itemize}
 \item $n$ is the number of qubit of the first register,
 \item $p\leq2^n$ the number of stored patterns,
 \item $q\leq p$ the number of stored patterns if the values of $t$ qubits are 
known
(i.e. $t$ qubits have been measured or are already disentangled to others or
the oracle acts on a subspace of $(n-t)$ qubits),
\item $c=\mathtt{ceil}(\log_2{q})$, i.e. the least integer greater or equal to
$\log_2{q}$,
\item $m\leq q$ the number of values $x$ for which $f(x)=1$,
\item $r=\mathtt{int}(\log_2{m})$ is the integer part of $\log_2{m}$.
\end{itemize}
\begin{remark}
 The goal of the NLA is to determine if a needed state exist in a register. The 
conditional gates are made to act only if this state exists in the register. 
Therefore gate $mathtt{S}$ can only computes memorised states even in case of 
completion problem.
\end{remark}

\section{Taking into account the quantum noise}
\label{sec:noise}

We will briefly analyze in this section how our QAM-NLSA evolves in the presence
of quantum noise. As the NLSA evolves qubit per qubit, we will consider only the
single qubit quantum noise channels as described in \cite{nielsen,cpwilliams}.
The quantum states to be considered will be the density operators instead of
state vectors.

\subsection{Single qubit quantum noise channels}

If $\rho_{in}=\ket{\psi}\bra{\psi}$ is the density matrix of the state
$\ket{\psi}$, the effect of the environment leads in the Kraus representation 
to
\begin{equation}
\rho_{out}=\mathcal{K}(\rho_{in})=\sum_iE_i\rho_{in}E^{\dag}_i,\,\,\sum_iE^{\dag
} _iE_i=\mathbb{I},
\end{equation}
where $E_i$ are the \emph{errors operators} or \emph{Kraus operators} which 
completely describe here the single qubit quantum noise channels briefly 
presented in the Table \ref{tab:1qbit-channels}.

\begin{table}[htbp]
\centering
\begin{tabular}{l|p{6cm}|p{5.5cm}}\hline\hline
Noisy channels & Description & Set of Kraus operators $\{E_i\}$ 
\\\hline\hline
Bit flip & Induced by dissipation, it flips the state $\ket{k}$ to 
$\ket{\bar{k}}, \,k=0,1$. & 
$\{\sqrt{1-\eta}\mathtt{X}^0,\sqrt{\eta}\mathtt{X}\}$\\\hline
Phase flip & Induced by decoherence, it flips the state $\ket{k}$ to 
$(-1)^k\ket{k},\,k=0,1$. & 
$\{\sqrt{1-\eta}\mathtt{Z}^0,\sqrt{\eta}\mathtt{Z}\}$ 
\\\hline
Bit-phase flip & It is a joint action of bit and phase flips. & 
$\{\sqrt{1-\eta}\mathtt{Y}^0,\sqrt{\eta}\mathtt{Y}\}$\\\hline
Amplitude damping & It transforms state $\ket{1}$ into state $\ket{0}$ but 
leaves state $\ket{0}$ unchanged. It should be viewed as energy dissipation. & 
$\{\ket{0}\bra{0}+\sqrt{1-\eta}\ket{1}\bra{1},\sqrt{\eta}\ket{0}\bra{1}\}$
\\\hline
Phase damping & It involves the loss of information about relatives phases in 
quantum state & 
$\{\ket{0}\bra{0}+\sqrt{1-\eta}\ket{1}\bra{1},\sqrt{\eta}\ket{1}\bra{1}\}$
\\\hline
Depolarizing channel & It transforms any state into a completed mixed state. 
&$\{\sqrt{1-\eta}\mathbb{I},\sqrt{\frac{\eta}{3}}\mathtt{X},\sqrt{\frac{\eta}{3}
}\mathtt{Y},\sqrt{\frac{\eta}{3}}\mathtt{Z}\}$\\\hline
\end{tabular}
\caption{Single qubit quantum noise channels. 
$\mathtt{X}_{i=0,1,2,3}=\mathbb{I},\,\mathtt{X},\,\mathtt{Y},\, \mathtt{Z}$ are 
Pauli matrices and $\eta\in[0,1]$ is the probability of a state to be affected 
by the noise. It should be noted that $\mathtt{X}_{i=1,2,3}^0=\mathbb{I}$ is 
not an error operator. Therefore, value $1-\eta$ associated to this operator 
represents the probability that no error occurs.}
\label{tab:1qbit-channels}
\end{table}

\subsection{Quantum associative memories with noise - bit flip model}
\label{sec:BitFlip}

We suppose that during the nonlinear evolution (NLE), step \ref{alg:Balgo13} to 
step \ref{alg:Ealgo13} of the Algorithm \ref{alg:algo1}, the quantum noise 
occurs with the probability $\eta\in[0,1]$ after the action of each gate of the
$\mathtt{NLE}$ gate. We assume that
\begin{itemize}
\item each gate operates before the error proceeds and it is the same error,
\item the first register is an $n$-qubit system and that the probability $\eta$ 
of a 
state to be affected by the noise is independent of the total number of network 
qubits,
\item errors are located at each time step in the network affecting 
$\ket{\ell}$ and $\ket{k}$.
\end{itemize}

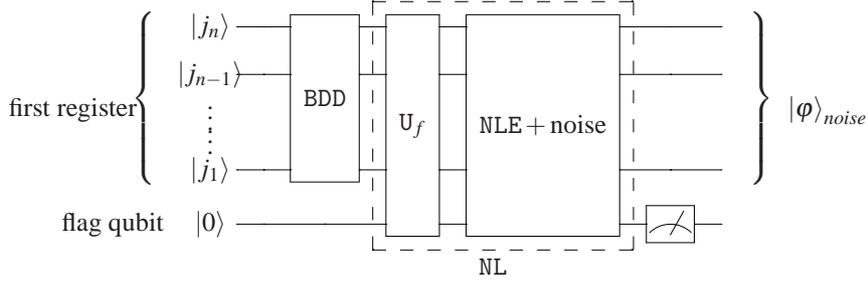
\begin{figure}[htbp]
\[ \Qcircuit @C=1.0em @R=.9em {
&&\ket{j_n}&&\qw&\multigate{4}{\mathtt{BDD}}&\multigate{5}{\mathtt{U}_f}
&\multigate{5}{\mathtt{NLE}+\text{noise}}&\qw &\qw&\\
&&\ket{j_{n-1}}&&\qw&\ghost{\mathtt{BDD}}&\ghost{\mathtt{U}_f}
&\ghost{\mathtt{NLE} +\text{noise}}&\qw&\qw&\\ 
&\lstick{\text{first register }\hspace{1.em}}
\gategroup{1}{1}{5}{1}{1em}{\{}&\vdots&& & & &  & & &\rstick{\hspace{1em} 
\ket{\varphi}_{noise}}\gategroup{1}{11}{5}{11}{1em}{\}}\\ 
& &\vdots& &  & & & & &
\\&&\ket{j_1}&&\qw&\ghost{\mathtt{BDD}}&\ghost{\mathtt{U}_f}
&\ghost{\mathtt{NLE}+\text{noise}}&\qw &\qw&\\
&\lstick{\text{flag qubit }}&\ket{0}&&\qw&\qw&\ghost{\mathtt{U}_f}
&\ghost{\mathtt{NLE}+\text{noise}}\gategroup{1}{7}{6}{8}{1em}{--}
&\meter&\qw&\\ 
&&&&&&\hspace{6em}\mathtt{NL}&&&
}
\]
 \caption{Schematic structure of the NLSA with quantum noise.}
 \label{fig:fig_algo3}
\end{figure}

Now, we analyse the effect of each quantum noise channels during evolution of
states (\ref{eq:equaO}). Thus, while we reduce the system to the two qubits
$\ket{\ell}$ and $\ket{k}$ we have
\begin{small}\begin{equation}
\label{phiOut}
\rho_{out}=\sum_iE_i\left(\mathtt{W}\otimes\mathtt{X}\left[
\sum_iE_i\left(\mathtt{NL}^+\left[\sum_iE_i\left(\mathtt{NL}^-\left[
\sum_iE_i\left(\mathtt{U}\rho_{in}
\mathtt{U}^\dag\right)E_i^\dag\right](\mathtt{NL}^-)^\dag\right)E_i^\dag\right
](\mathtt{ NL}^+)^\dag\right)E_i^\dag\right] \mathtt{W} ^\dag\otimes\mathtt{X}
^\dag\right)E_i^\dag.
\end{equation}              \end{small}
Because for each step of NLE, we look a two qubits system, then errors operators
to apply will be a tensor product of operators of single quantum noise models.

For the bit flip model, the set of operators where $r$ index the qubit of first
register and $f$ index the flag qubit is given below by the tensor product
$\{\mathbb{I}_r,\mathtt{X}_r\}\otimes\{\mathbb{I}_f,\mathtt{X}_f\}$:
\begin{equation}
 \begin{split}
  E_1&=\sqrt{1-\eta_r}\sqrt{1-\eta_f}\mathbb{I}_r\otimes\mathbb{I}_f\\
  E_2&=\sqrt{1-\eta_r}\sqrt{\eta_f}\mathbb{I}_r\otimes\mathtt{X}_f\\
  E_3&=\sqrt{\eta_r}\sqrt{1-\eta_f}\mathtt{X}_r\otimes\mathbb{I}_f\\
  E_4&=\sqrt{\eta_r}\sqrt{\eta_f}\mathtt{X}_r\otimes\mathtt{X}_f.
 \end{split}
\end{equation}
Remark that according to equation (\ref{phiOut}), $\rho_{out}$ is the sum of 
$256$ matrix for bit flip model. 

Due to entanglement the system must be observed completely. So, without 
 the normalization constant, the input matrix is
\begin{equation}
\begin{split}
 \rho_{in}&=\ket{\psi}\bra{\psi}\\
            &=\left[\sum_{\substack{j_nj_{n-1}\hdots j_2=0\\
j_nj_{n-1}\hdots j_2\neq i_ni_{n-1}\hdots i_2}}^1\ket{j_nj_{n-1}\hdots j_20}
\ket{0}+\ket{j_nj_{n-1}\hdots j_21}\ket{0}+\ket{i_ni_{n-1}\hdots(1-i_1)}\ket{0}
+\ket{i_ni_{n-1}\hdots i_1}\ket{1}\right]\\&\left[\sum_{\substack{j_nj_{n-1}
\hdots j_2=0\\j_nj_{n-1}\hdots j_2\neq i_ni_{n-1}\hdots i_2}}^1\bra{j_nj_{n-1} 
\hdots j_20}\bra{0}+\bra{j_nj_{n-1}\hdots j_21}\bra{0}+\bra{i_ni_{n-1} 
\hdots(1-i_1)} \bra{0} +\bra{i_ni_{n-1}\hdots i_1}\bra{1}\right],
 \end{split}
\end{equation}
while the sought output density matrix is
\begin{equation}
 \Omega=\left(\left[\sum_{j_nj_{n-1}\hdots j_1=0}^1\ket{j_nj_{n-1}\hdots j_1}
\right]\left[\sum_{j_nj_{n-1}\hdots j_1=0}^1\bra{j_nj_{n-1}\hdots j_1}
\right]\right)\otimes\ket{1}\bra{1}.
\end{equation}

To evaluate the influence of the quantum noise on the effectiveness of the 
algorithm, we will compute the \emph{fidelity} between the sought output and 
the obtained output. Let $\sigma$ be the density matrix of the sought output. 
The fidelity then is
\begin{equation}
 \mathcal{F}_1(\sigma,\Gamma)=tr\sqrt{\sqrt{\sigma}\Gamma\sqrt{\sigma}},
 \label{eq:equaFidel}
\end{equation}
with $0\leq\mathcal{F}_1(\sigma,\Gamma)\leq1$. $\mathcal{F}_1(\sigma,\Gamma)=0$ 
if $\sigma$ and $\Gamma$ are orthogonal and $\mathcal{F}_1(\sigma,\Gamma)=1$ 
if $\sigma = \Gamma$.

If we only focus on the effectiveness of the quantum noise on the NLE, that is 
before the retrieving process, we can consider that the sought output is a pure 
state $\ket{\psi}=\sum_x\ket{x} =\ket{\phi_1}\otimes\ket{\phi_2}\dots\otimes 
\ket{\phi_n},\,\ket{\phi_j}=\alpha_j\ket{0}+\beta_j\ket{1}$ ($|\alpha_j|^2+ 
|\beta_j|^2=1$), and then the equation (\ref{eq:equaFidel}) can be written as
\begin{equation}
 \mathcal{F}_0(\ket{\psi},\rho_{out})=\sqrt{\bra{\psi}\rho_{out}\ket{\psi}}.
 \label{eq:equaFidel1}
\end{equation}

\subsection{Simulation}
We suppose that
\begin{equation}\label{eq:Mnl-}
\begin{split}
 \mathtt{NL}^-&=\mathtt{CX}^1_{fr}\mathtt{CW}^0_{rf}\mathtt{CW}^1_{rf}\\
	      &=\frac{1}{\sqrt{2}}\begin{pmatrix}
	                    1&1&0&0\\
	                    0&0&1&-1\\
	                    0&0&1&1\\
	                    1&-1&0&0
\end{pmatrix},
 \end{split}
\end{equation}
and according to equations (\ref{eq:Mnl-}), we can choose 
$\mathtt{M}=\mathtt{W}$. We also consider that $\eta_r=\eta_f=\epsilon$, 
according to the fact the error operator is applied on states $\ket{\ell}$ and 
$\ket{k}$ simultaneously (i.e., error arises on this qubits at the same time),
thus we can consider that the probability of error is identical for both 
qubits. The set of operators is now

\begin{equation}
 \begin{split}
  E_1&=(1-\epsilon)\mathbb{I}_r\otimes\mathbb{I}_f\\
  E_2&=\sqrt{\epsilon(1-\epsilon)}\mathbb{I}_r\otimes\mathtt{X}_f\\
  E_3&=\sqrt{\epsilon(1-\epsilon)}\mathtt{X}_r\otimes\mathbb{I}_f\\
  E_4&=\epsilon\mathtt{X}_r\otimes\mathtt{X}_f.
 \end{split}
\end{equation}

Because it is the flag qubit which is observed, we only extract its density 
matrix $\rho_{out}$ to evaluate fidelity.

\subsubsection{Case where the sought state does not exist in the first register}
Here the sought output of the flag qubit is $\ket{0}$.
\begin{subequations}
 If the first register has $n$ qubits,
 \begin{equation}
  \rho_{out}=\begin{pmatrix}
            1+\frac{1}{2}[(1-2\epsilon)^{3n}-1] & 0\\
            0 & -\frac{1}{2}[(1-2\epsilon)^{3n}-1]
           \end{pmatrix},
 \end{equation}
then
 \begin{equation}
 \mathcal{F}_0(\ket{0},\rho_{out})=\sqrt{1+\frac{1}{2}[(1-2\epsilon)^{3n}-1]}.
 \label{eq:equaFidelFlag0}
\end{equation}
\end{subequations}
Indeed,
\begin{itemize}
 \item if the first register has only one qubit,
 \begin{equation}
  \rho_{out}=\begin{pmatrix}
            1+\frac{1}{2}[(1-2\epsilon)^3-1] & 0\\
            0 & -\frac{1}{2}[(1-2\epsilon)^3-1]
           \end{pmatrix};
 \end{equation}
\item if the first register has three qubits,
 \begin{equation}
  \rho_{out}=\begin{pmatrix}
            1+\frac{1}{2}[(1-2\epsilon)^9-1] & 0\\
            0 & -\frac{1}{2}[(1-2\epsilon)^9-1]
           \end{pmatrix}.
 \end{equation}
\end{itemize}

\begin{figure}[htbp]
\centering
 \leavevmode
 \subfloat[If $n$ is odd]{
 \includegraphics[scale=0.75]{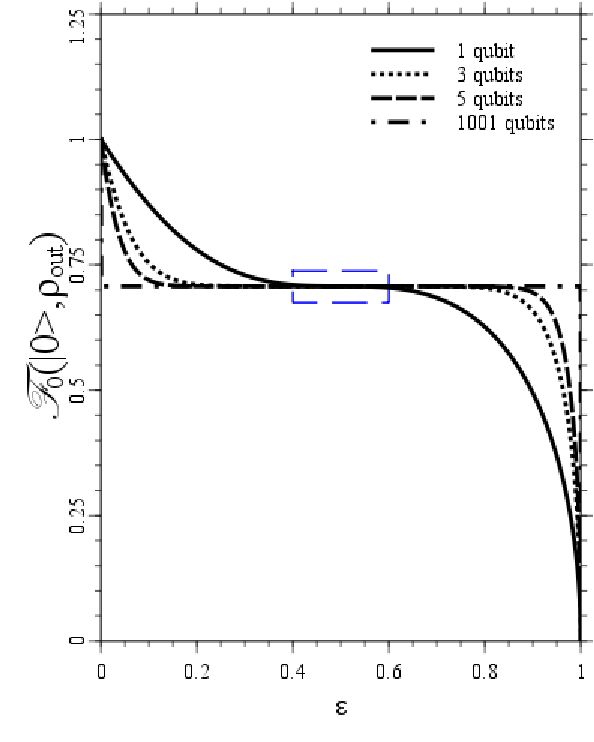}\label{fig:fidelitya}}
 \subfloat[If $n$ is even]{
 \includegraphics[scale=0.75]{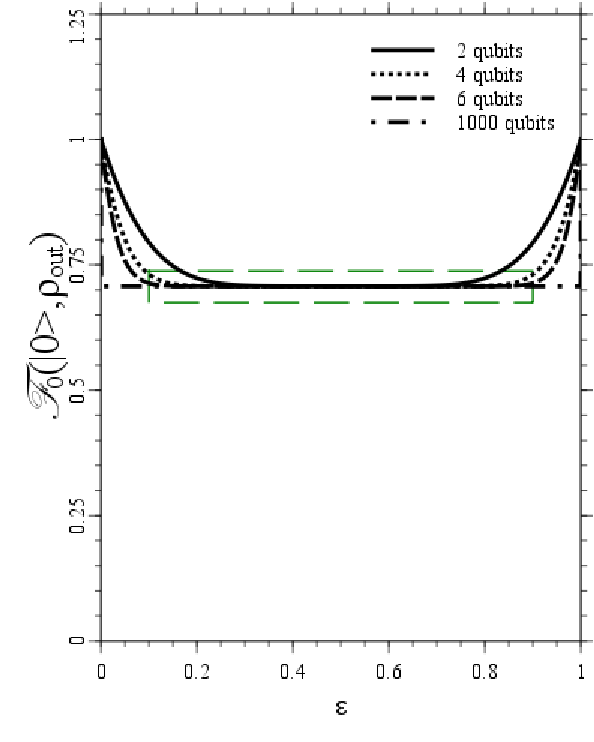}\label{fig:fidelityb}}
 \caption{Evolution of the fidelity $\mathcal{F}_0(\ket{0},\rho_{out})$ at the 
end of NLE steps, in the case where the sought state does not exist in the 
first register, (a) for $1$, $3$, $5$ and $1001$ qubits and (b) for $2$, $4$, 
$6$ and $1000$ qubits. The dashed blue and green rectangles enlighten the 
stability area.}
 \label{fig:fidelity}
 \end{figure}

As we shall see on Fig. \ref{fig:fidelity} and according to equation
(\ref{eq:equaFidelFlag0}), while $\epsilon<0.5$ the fidelity is upper than
$0.7$. But it decreases for $\epsilon\geq0.5$ if the $n$ is an odd number, and
increases if $n$ is an even one. For $n$ as an odd number (see Fig.
\ref{fig:fidelitya}, one qubit), the area between $0.4$ and $0.6$ can be viewed
as a stability area, i.e., the area where fidelity is maintained around $0.7$.
For $n$ as an even number (see Fig. \ref{fig:fidelityb}, six qubits), this
stability area grows with the number of qubit. Therefore, if the sought state
does not exist in the first register, the QAM-NLSA is more resistant to noise
when the register has an even number of qubit.

\begin{figure}[htbp]
\centering
 \includegraphics[scale=0.75]{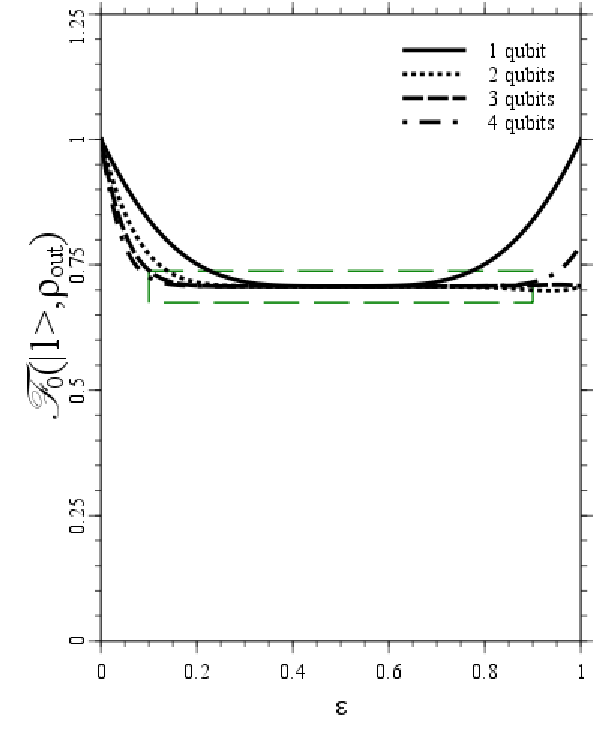}
 \caption{Evolution of the fidelity $\mathcal{F}_0(\ket{1},\rho_{out})$ 
at the end of NLE steps for $1$, $2$, $3$ and $4$ qubits in the case where 
the sought state exists in the first register.}
 \label{fig:fidelity1}
 \end{figure}

\subsubsection{Case where the sought state exists in the first register}

Here the sought output of the flag qubit is $\ket{1}$.
\begin{subequations}
If the first register has $n$ ($n>1$) qubits
 \begin{equation}
  \rho_{out}=\begin{pmatrix}
            -\frac{1}{2}\frac{f(\epsilon^{4n})}{2^{2n-3}} & 0\\
            0 & 1+\frac{1}{2}\frac{f(\epsilon^{4n})}{2^{2n-3}}
           \end{pmatrix},
 \end{equation}
where $f$ is a polynomial function which grows with $\epsilon^{4n}$. Then,
\begin{equation}
 \mathcal{F}_0(\ket{1},\rho_{out})=\sqrt{1+\frac{1}{2}[(1-2\epsilon)^4-1]}, 
\text{ for }n=1,
 \label{eq:equaFidelFlag1}
\end{equation}
or
\begin{equation} 
\mathcal{F}_0(\ket{1},\rho_{out})=\sqrt{1+\frac{1}{2}\frac{f(\epsilon^{4n})} 
{2^{2n-3}}}\text{ for }n>1.
 \label{eq:equaFidelFlag12}
\end{equation}
\end{subequations}
Indeed,
\begin{itemize}
 \item if the first register has only one qubit,
 \begin{equation}
  \rho_{out}=\begin{pmatrix}
            -\frac{1}{2}[(1-2\epsilon)^4-1] & 0\\
            0 & 1+\frac{1}{2}[(1-2\epsilon)^4-1]
           \end{pmatrix};
 \end{equation}
\item if the first register has two qubits,
 \begin{equation}
  \rho_{out}=\begin{pmatrix}
            -\frac{\frac{1}{2}[((1-2\epsilon)^8-1)+((1-2\epsilon)^7-1)]}{2} 
& 0\\
            0 & 1+\frac{\frac{1}{2}[((1-2\epsilon)^8-1)+((1-2\epsilon)^7-1)]}{2}
           \end{pmatrix};
 \end{equation}
 \item if the first register has four qubits,
 \begin{equation}
  \begin{pmatrix}
-\frac{
\frac{1}{2}\begin{bmatrix}
((1-2\epsilon)^{16}-1)+5((1-2\epsilon)^{15}-1)\\+15((1-2\epsilon)^{14}-1) 
+13((1-2\epsilon)^{13}-1)\\+0((1-2\epsilon)^{12}-1)-6((1-2\epsilon)^{11}-1)\\+
5((1-2\epsilon)^{10}
-1)-((1-2\epsilon)^9-1)\\-((1-2\epsilon)^8-1)+((1-2\epsilon)^7-1)\end{bmatrix}} 
{2^5} & 0\\
   0 & 1+\frac{
\frac{1}{2}\begin{bmatrix}
((1-2\epsilon)^{16}-1)+5((1-2\epsilon)^{15}-1)\\+15((1-2\epsilon)^{14}-1) 
+13((1-2\epsilon)^{13}-1)\\+0((1-2\epsilon)^{12}-1)-6((1-2\epsilon)^{11}-1)\\+
5((1-2\epsilon)^{10}
-1)-((1-2\epsilon)^9-1)\\-((1-2\epsilon)^8-1)+((1-2\epsilon)^7-1)\end{bmatrix}} 
{2^5}
  \end{pmatrix}.
   \end{equation}
\end{itemize}

As we see on Fig. \ref{fig:fidelity1} and according to equation 
(\ref{eq:equaFidelFlag12}), whatever the value of $\epsilon$ the fidelity is 
greater than $0.7$. In other words, if the sought state exists in the first 
register, the QAM-NLSA is resistant to noise whatever the number of qubit. We 
also see that the stability area enlightened by the dashed green rectangle on 
Fig. \ref{fig:fidelity1} grows with the number of qubit. That stability area  
is the same as 
those shown on Fig. \ref{fig:fidelity}.

From the above simulations, it appears that QAM-NLSA is affected by the noise 
during its implementation. In the particular case of bit flip channel, the 
fidelity between the unaffected and affected systems is about $70\%$ and this 
value does not change even if the number of qubit in the first register grows.

\section{Conclusion}
\label{sec:Concl}

We have proposed a model of the QAM-NLSA similar to that of Rigui
\etal\cite{zhou2012}. However, the model we propose differs with the possibility
to retrieve one of the sought states in multi-values retrieving when a measure
on the first register is done.

Firstly, we have described the NLSA put forth by Abrams and Lloyd in
\cite{Abrams1998} with notations that overcome some ambiguities due to the
notations of Rigui \etal and Czachor \cite{Czachor98} and by summarizing each
step of the nonlinear evolution with an equivalent circuit. A good general form
of the unitary matrix $\mathtt{NL}^+$ which acts on the generic flag qubit
$\alpha\ket{0}+\beta\ket{1}$ was given thereby correcting the wrong one given by
Rigui \etal. Secondly, we have described our model of the Quantum Associative
Neural Network where we have introduced a $(2^{n+1})\times(2^{n+1})$ conditional
operator $\mathtt{CS}$ which maps the first register to the sought state
$\ket{x}$ when the flag qubit is $\ket{1}$, where $n$ is the number of qubit of
the first register. If $n$ is the number of qubit of the first register,
$p\leq2^n$ the number of stored patterns, $q\leq p$ the number of stored
patterns if the values of $t$ qubits are known (i.e., $t$ qubits have been
measured or have already been disentangled to others or the oracle acts on a
subspace of $(n-t)$ qubits), $m\leq q$ the number of values $x$ for which
$f(x)=1$, $c=\mathtt{ceil}(\log_2 {q})$ the least integer greater or equal to
$\log_2{q}$, and $r=\mathtt{int} (\log_2 {m})$ the integer part of $\log_2{m}$,
then the time complexity of our algorithm is $\mathcal{O}(c-r)$. It is better
than Grover's algorithm and its modified forms which need $\mathcal{O}
(\sqrt{\frac{2^n} {m}})$ steps when they are used as the retrieval algorithm. An
example to illustrate the results given by our analysis was done. It is
noteworthy that our algorithm also allows to measure the flag qubit to erase any
ambiguity on the result given by a measurement on the first register. This 
possibility is 
introduce by the use of two conditional gates who do not affect the flag qubit 
after nonlinear evolution.

Finally, we have briefly analysed the influence of the quantum noise namely bit
flip channel on our model of the QAM-NLSA. We found that the bit flip channel
leaves the QAM unaffected fully at $70\%$ if the sought state is present in the
first register or if the register has an even number of qubit when the sought
state does not exist. However, when the first register has an odd number of
qubit and the sought state does not exist, the bit flip channel is extremely
destructive when the probability $\epsilon>0.5$.

Further work will be undertaken in order study in details the influence of 
quantum noise related to the quantum network construction through errors 
characterizing the qubit time evolution and gate application in both the first 
register and the flag qubit.

\section{Acknowledgments}
 Authors are very grateful to Dr D.E. Houpa Danga for useful discussions and
 remarks. We are thankful to M. A.L. Kamga for proofreading the work.

\bibliographystyle{unsrt}

\end{document}